# Roadmap and direction towards high performance $MoS_2$ hydrogen evolution catalysts

Yang Cao
Department of Energy and Resources Engineering, College of Engineering, Peking University, Beijing, 100871 P. R. China.

**Abstract**
$MoS_2$ is a typical 2D transition metal dichalcogenide, and have exhibit superior electrocatalytic HER activity. The $MoS_2$'s HER catalysis was developed 15 years ago, various nanofabrication strategies are applied to boost the performance. It is promising that $MoS_2$ would take place of Pt in HER catalysis. Various active catalytic sites, including edge, vacant, basal plane, *etc*. are developed and the catalytic performance were compared. Hybrid composition were developed, like modification with atoms, clusters, loading on substrates were developed. We make a summary on HER mechanisms varies with active sites and operation solutions. $MoS_2$ synthesis, characterization, HER performance, mechanisms, to make a holistic understanding on the interplay between the structure, chemistry, HER performance, and mechanism. It is believed that the review will help researchers to get a better understanding on $MoS_2$'s superior HER performance, and provide a wealth of catalyst tool box to promote next-generation catalysts development.

**Main text**

Hydrogen energy is perceived as the most promising clean energy without any carbon emission like $CO_2$. For the greenhouse effect is becoming increasingly serious issue caused by large application of fossil fuel, hydrogen fuel is an innovative solution and show great potential in resolute carbon emission. Electrocatalytic water splitting is the best achievable approach nowadays for achieving large amount of $H_2$.

$MoS_2$ are facile, stable, non-toxic, affordable materials with reasonable price, and have exhibit its superior potential in catalysis, sensing, opt electrochemical, environmental relate application[1], *etc*. Modification of $MoS_2$ materials has been widely applied for adjusting and formulate the relative performance in certain application field. $MoS_2$ is a typical transition metal dichalcogenide[2] (TMD) compound with a two-dimensional S-Mo-S tri-atom layer structure.

$MoS_2$ is predicted to be a promising substitute catalyst for platinum 15 years ago[3]. While pristine $MoS_2$ shows low electrocatalysis performance, via vacancies engineering, phase engineering, heterojunction engineering, hetero-atom doping strategies, the catalytic active sites quantities are elevated and catalysis performance is magnitude improved.

We summarize on $MoS_2$ synthesis, characterization, HER performance, mechanisms, to make a holistic understanding on the interplay between the structure, chemistry, HER performance, and mechanism, which will give directions to elevate other electrochemical reaction by $MoS_2$.

**Structural, fundamental physical and chemical properties of $MoS_2$:**

$MoS_2$ has been applied as lubricant materials at industrial level in alleviate worn out phenomenon[4-5]. Generally, two-dimensional $MoS_2$ exist 4 main crystal structure, which are 1H, 1T, 2H, 3R[6]. Among them, 1H $MoS_2$ is the most stable phase. $MoS_2$ is composed with S-Mo-S trilayer atoms, whereas the layers are weakly coupled and the interlayer space is around 3 Å between the adjacent two layers. 2H, 1T phase of $MoS_2$ are composed with hexagonally Mo atoms or tetragonal Mo atoms, and they are semiconducting or metallic, respectively. Which is fascinating to chemist and physicist is the adjustable structure and correlate unique electronic properties, catalytic activities. $MoS_2$ exhibit a wealthy of quantum physics like quantum spin Hall insulator and superconducting states[7] (Figure 1). It has been posited that the favored crystal phase can be easily switched by regulate the electron filling of d orbitals[7-8].

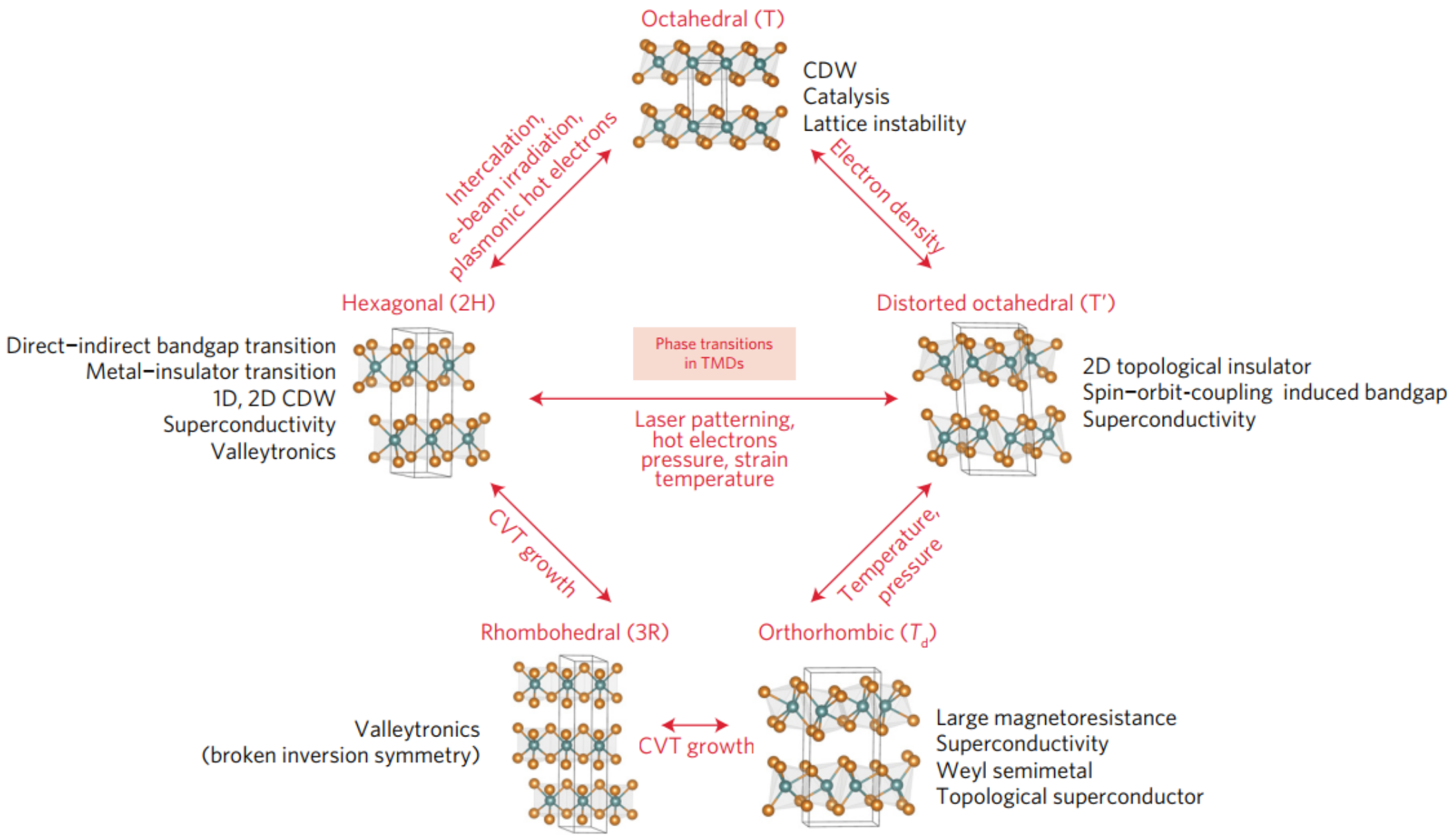

**Figure 1.** Transition metal dichalcogenides compound phase transition and correlation between different phase and physics[7] (Copyright 2017, Nature Publishing Group).

$MoS_2$'s structure is closely related to its electronic structure. As depicted in Figure 2a, pristine $MoS_2$'s structure trigonal prismatic (Figure 2a, b), but 1T phase $MoS_2$ is octahedral[7] (Figure 2a, c). The natural bulky $MoS_2$ exhibit as semiconducting 2H phase with an indirect band gap of 1.3 eV. when the thickness is reduced to around 100 nm, the band energy is elevated to 1.6 eV caused by quantum confinement effect. When the thickness is reduced to monolayer, the band gap is transformed from indirect to direct with the predict theoretical band gap ~1.9 eV. The disordered 1T′ phase (partly metallic 1 T phase) is still semiconductor with a band gap close to zero (0.1 eV)[9]. When $MoS_2$ is 1T phase, $MoS_2$ is totally changed to metallic without any band-gap energy. As depicted in Figure 3, when convert from 2H towards 1T phase, $MoS_2$ lattice structure and electronic structure evolved along with electron injection, electron rearrangement, charge loss[10]. And the resultant 1T $MoS_2$ can reversed to pristine 2H $MoS_2$ by charge rearrangement. Suenaga *et. al.* traced the phase conversion from 2H to 1T using TEM and find that process undergo atomic glide (Figure 3b) [11], and α intermediate phase were identified under $MoS_2$ monolayer. During phase conversion, some other metastable distort phase were also formed (1T′, 1T″, 1T‴)[12]. In these metastable phases, zigzag Mo-Mo chains and dimerization Mo (1T″) or trimerization Mo (1T‴) atoms with closer Mo-Mo distance were found[13].

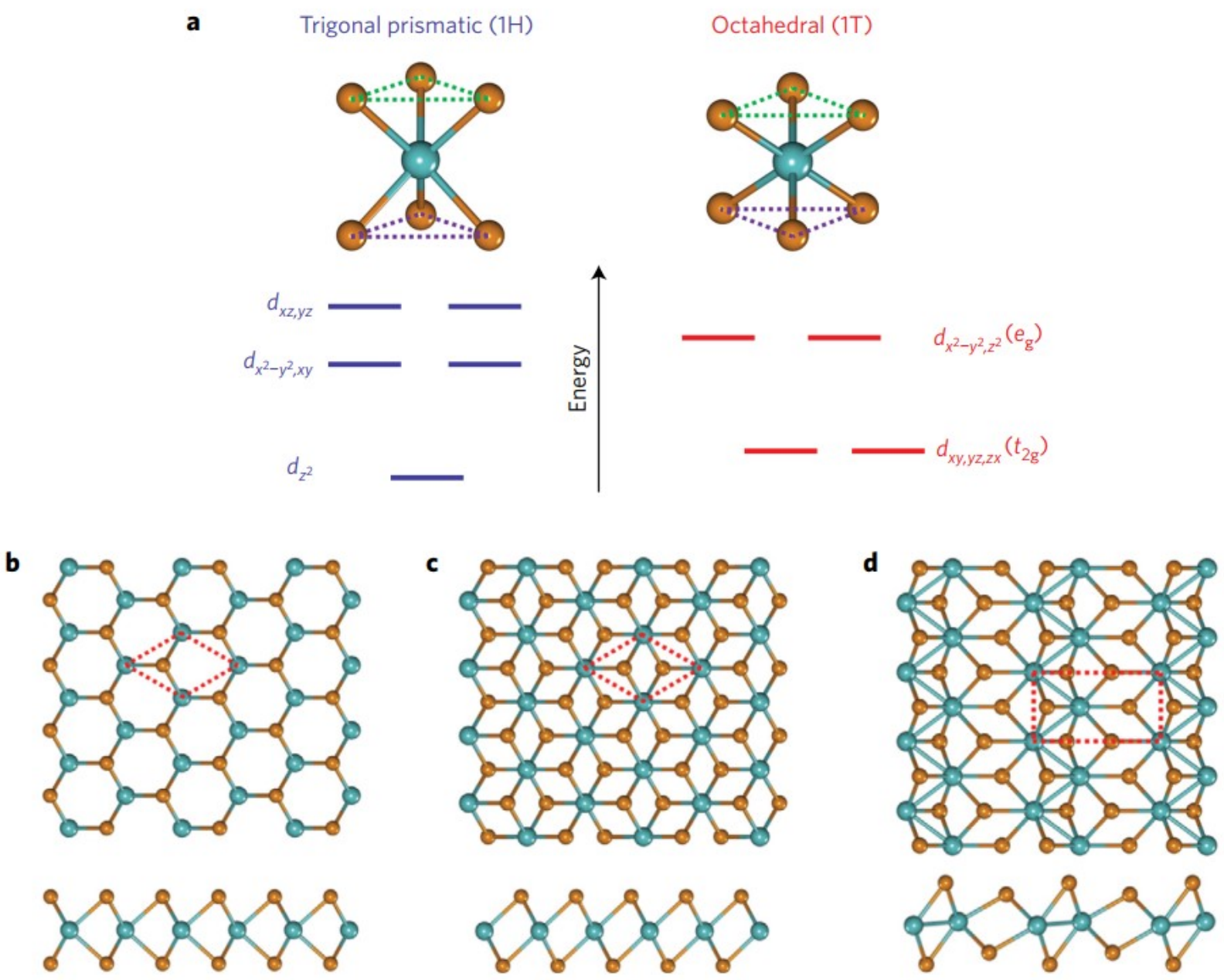

**Figure 2.** Structures of single-layer TMDs. **a**, Schematic images of 1H and 1T lattice symmetries and energy levels of *d*-orbital electrons induced by the crystal field. **b**–**d**, Top and side structures of 1H (**b**), 1T (**c**) and distorted 1T or 1T' (**d**). Blue spheres represent transition metals, and orange spheres denote chalcogen elements[7] (Adapted with permission from ref. 7. Copyright 2017, Nature Publishing Group).

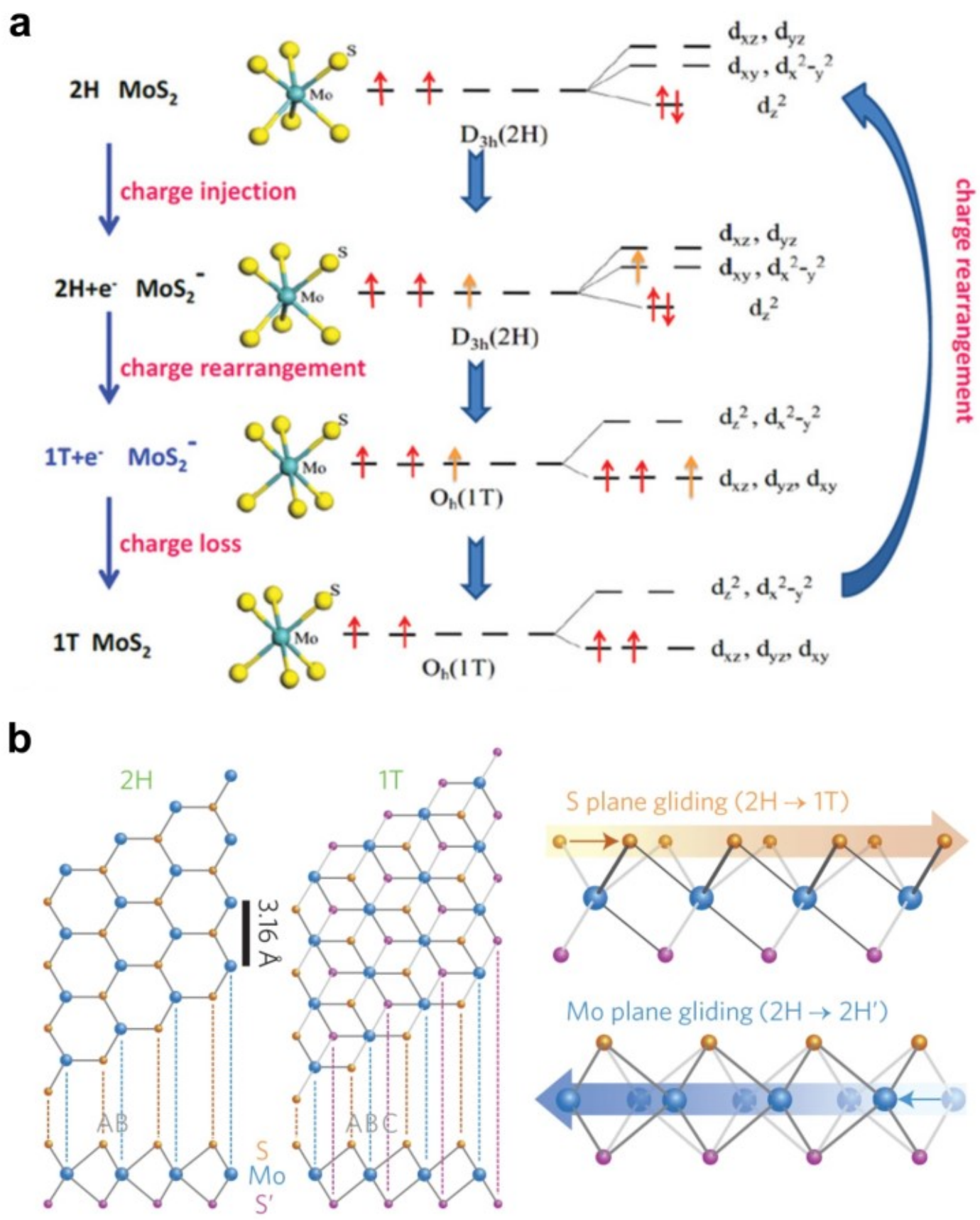

**Figure 3.** Phase transition between 1T and 2H mechanism based on the crystal field theory[11]. Schematic models of single-layered $MoS_2$ with 2H (a) (Adapted with permission from ref. 10. Copyright 2019, Wiley-VCH Verlag GmbH & Co. KGaA, Weinheim) and 1T (b) phases in basal plane and cross-section views (Adapted with permission from ref. 11. Copyright 2014, Nature Publishing Group)

**Determination on $MoS_2$'s structure and the relative characterization:**

Determination of the morphology, structure and phase are focal points for $MoS_2$ materials, correlate the catalytic performance with the structure, identify the catalytic active site. The structural represent direct relationship with $MoS_2$'s HER catalytic performance, structure examination is a vital issue for understanding and design stable $MoS_2$ catalyst with high performance.

**Powdered XRD characterization:**

By examine the XRD data, the crystallization behavior, phase constitution, crystal lattice information of the $MoS_2$ can be arrived. As depict in Figure 4a, 2H and 1T' would be distinguished for the (002) peak position. Some displaced 1 T phase (1T″, 1T‴) that difficult to identify as well as uncommon 3R phase can be indexed with the XRD results (Figure 4b, c)[14-17]. What's more, as depicted in Figure 5, (002) lattice plane (around 8~13°) is often applied to estimate $MoS_2$'s the interlayer distance by Bragg's law[18].

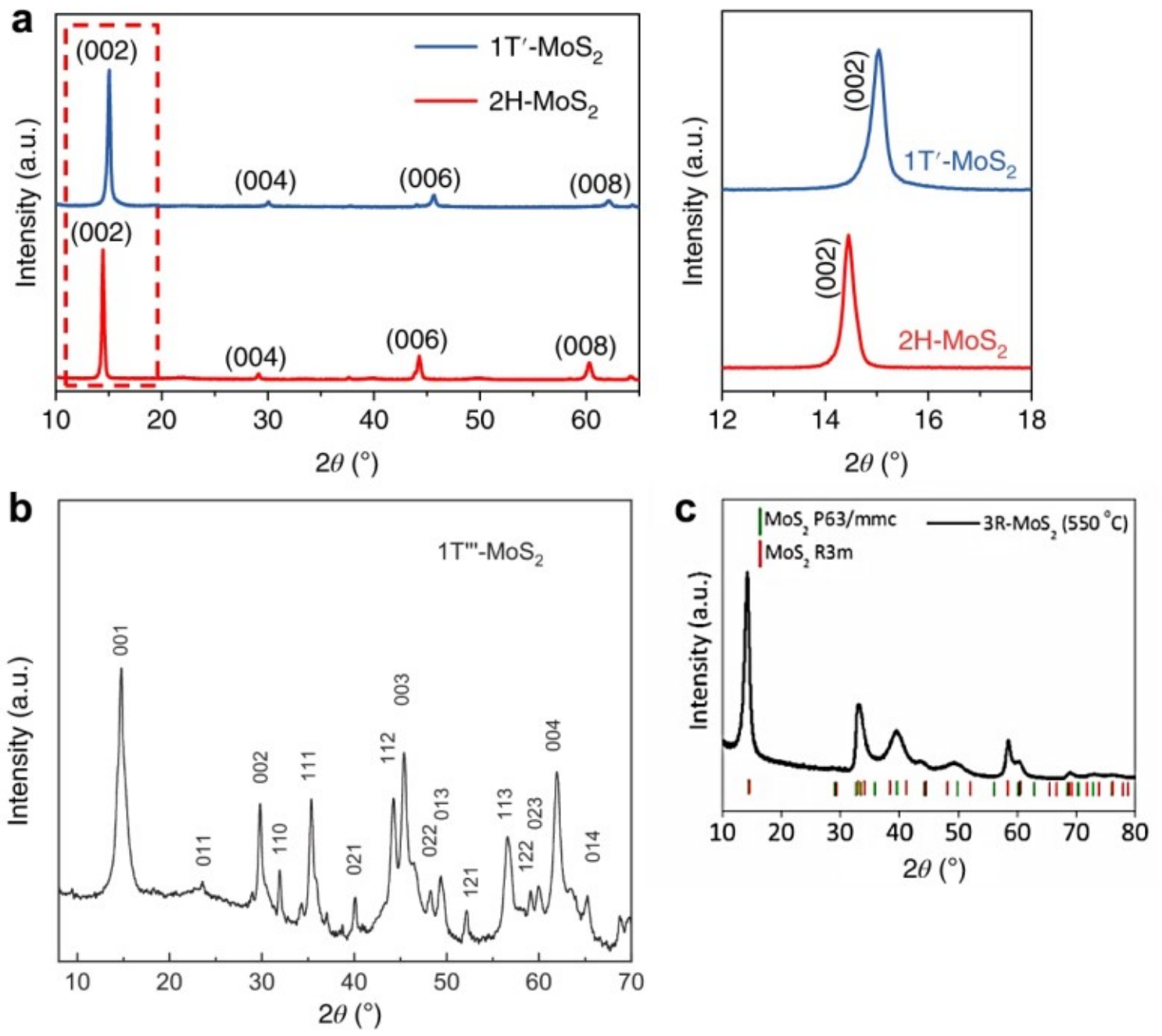

**Figure 4.** XRD patterns of (a) 1T′-$MoS_2$ crystals and 2H-$MoS_2$ crystals[16] (Adapted with permission from ref. 16. Copyright 2018, Nature Publishing Group) (b) 1T‴-$MoS_2$[15] (Adapted with permission from ref. 15. Copyright 2019, American Physical Society.) (c) 3R phase $MoS_2$[17] (Adapted with permission from ref. 17. Copyright 2019, Royal Chemical Society.)

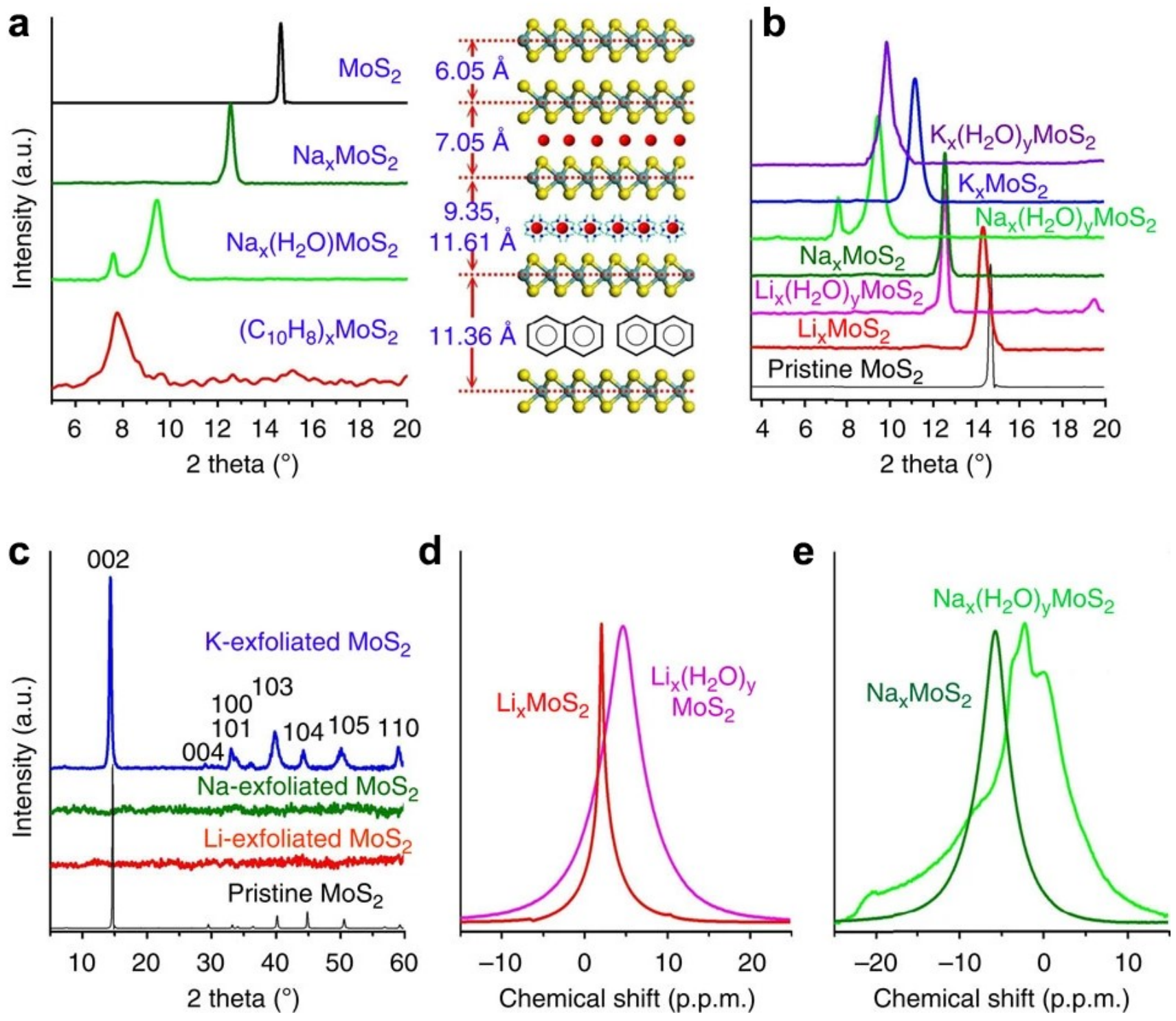

**Figure 5.** (a) XRD pattern and schematic of pristine $MoS_2$, Na-intercalated $MoS_2$, after exposure of Na-intercalated $MoS_2$ to the ambient for 3 days, exfoliated-and-restacked naphthalene-intercalated $MoS_2$. (b) Li-, Na- and K-intercalated $MoS_2$, after exposure of intercalated sample to the ambient for 3 days. (c) Li-, Na- and K-exfoliated $MoS_2$ without any annealing. (d) Solid-state $^7$Li NMR spectra of $Li_xMoS_2$ and $Li_x(H_2O)_yMoS_2$, (e) Solid-state $^{23}$Na NMR spectra of $Na_xMoS_2$ and

$Na_x(H_2O)_yMoS_2$[18]. (Adapted with permission from ref. 18. Copyright 2014, Nature Publishing Group)

**UV-Vis spectra and photo luminance characterization:**

It has been determined that 2H $MoS_2$'s absorption at around 600 and 670 nm (Figure 6a) are attribute to energy splitting from valance band and spin-orbit coupling[19]. 1T $MoS_2$ do not show any absorption ~600 nm for it belongs to metallic phase[20]. UV-Vis spectra can reflect on the structure of $MoS_2$, *i.e.* in Figure 6b, $MoS_2$ quantum dot shows strong absorption at ~300 nm, which is distinct from nanosheet or bulk $MoS_2$[21]. $MoS_2$ nanosheet and $MoS_2$ bulk display much stronger absorption in visible light region.

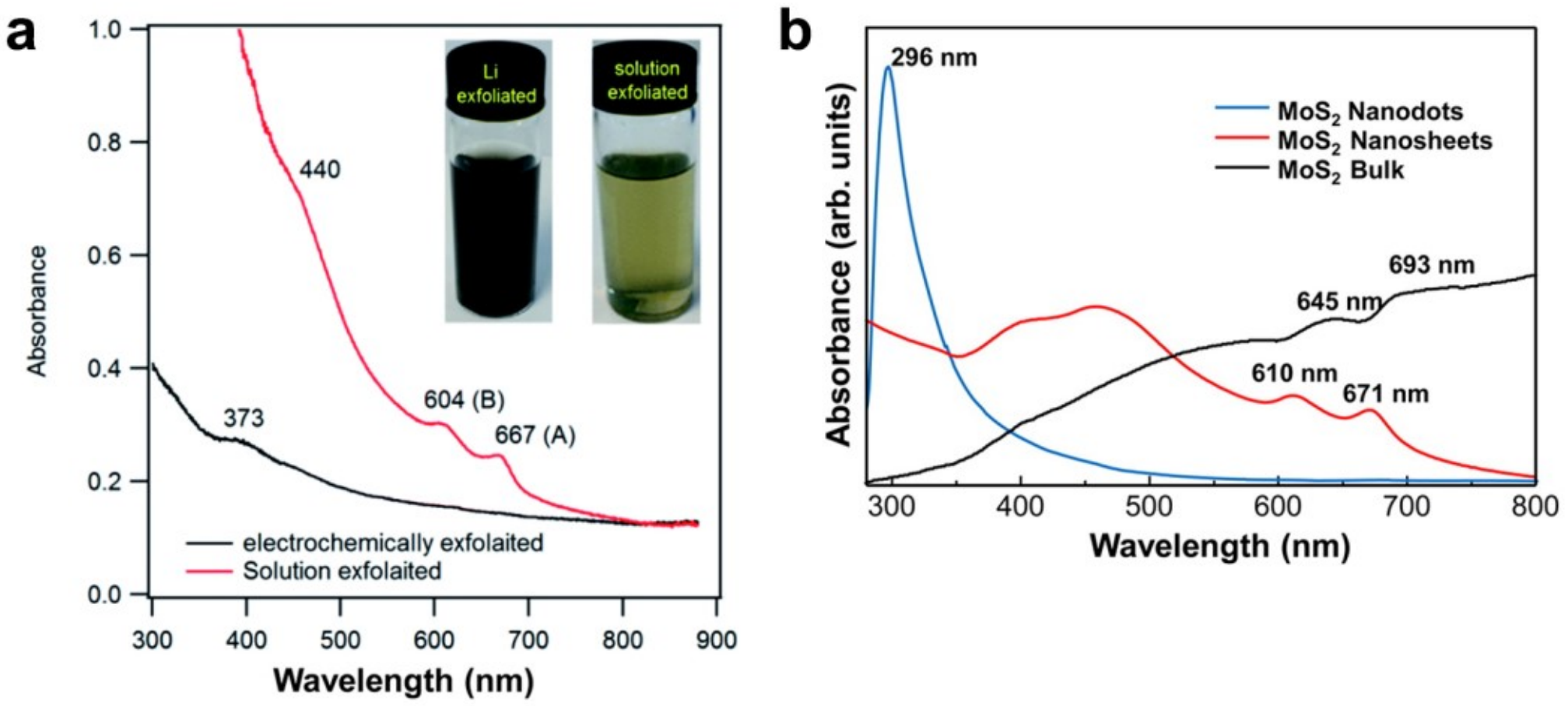

**Figure 6.** Optical absorption spectrum of $MoS_2$ (a) $MoS_2$ (Li exfoliated $MoS_2$: metallic phase, solution exfoliated $MoS_2$: semiconducting phase) dispersion in the mixture of water and isopropanol (1:1)[20]. (Adapted with permission from ref 20, Copyright 2017, Royal Chemical Society.) (b) $MoS_2$ nanodots, nanosheets, and bulk materials[21]. (Adapted with permission from ref 21. Copyright 2016, American Chemical Society.)

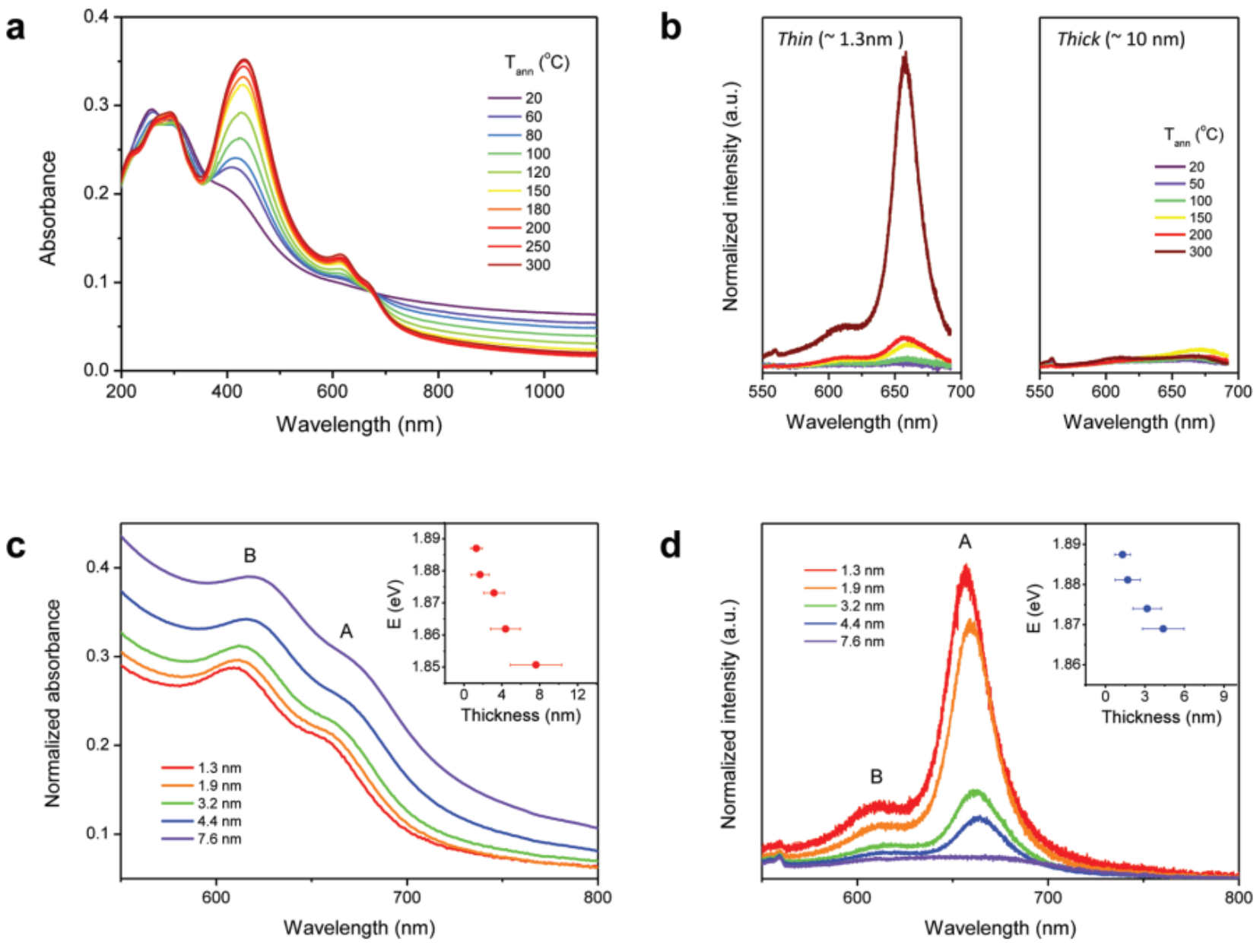

**Figure 7** (a) Absorption and (b) photoluminescence spectra of $MoS_2$ thin films annealed at various temperatures. (c) Absorption and (d) photoluminescence spectra of $MoS_2$ thin films with average thicknesses ranging from 1.3 to 7.6 nm[22]. (Adapted with permission from ref 22. Copyright 2011, American Chemical Society)

The photo luminance intensity and excitation position are sensitive to the structure and thickness of $MoS_2$, few layered $MoS_2$ has been exfoliated and studied[22]. As depicted in Figure 7, when the thickness of $MoS_2$ is increased from 1.3 nm to 7.6 nm, visible light absorption is elevated, while photoluminescence is decayed, when $MoS_2$ is reached to 7.6 nm, the photoluminescence vanished.

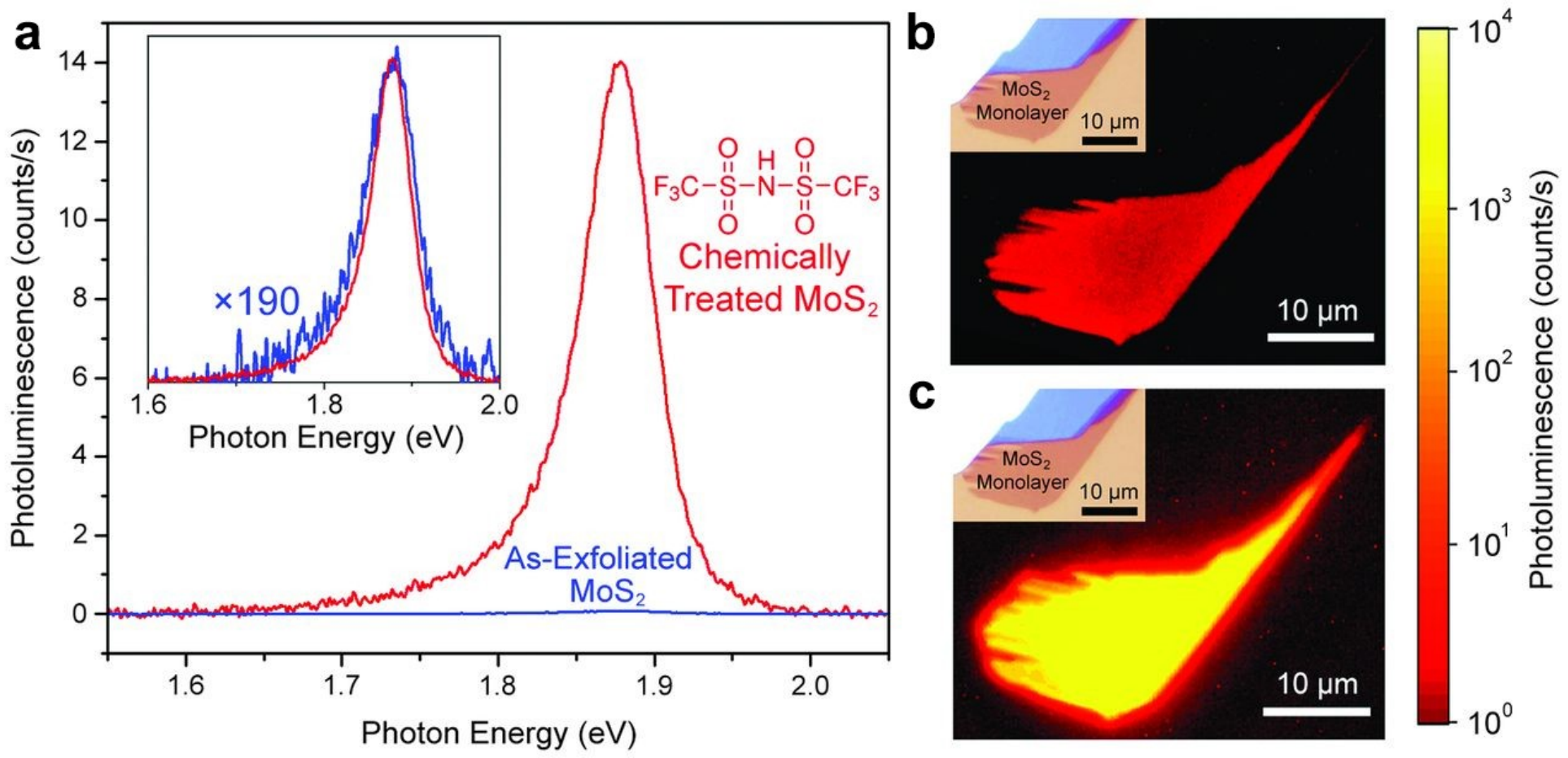

**Figure 8** (a) PL spectrum for both the as-exfoliated and TFSI-treated $MoS_2$ monolayers measured at an incident power of $1 \times 10^{-2}$ W $cm^{-2}$. The inset shows normalized spectra. (b) PL images of a $MoS_2$ monolayer before (b) and after treatment (c). Insets show optical micrographs[23]. (Adapted with permission from ref 23. Copyright 2015, American Association for the Advancement of Science)

Another study shows that the $MoS_2$'s poor photo luminance can be fine-tuned by surface organic molecule modification strategies. As depicted in Figure 8, the pristine photo luminance of $MoS_2$ display poor luminescence quantum yield (0.01 % to 6 % in the available studies) for large amount of defects, whereas nearly 100 % quantum yield were achieved by surface modification with TFSI molecules[23]. What's more, chemical doping ($F_4TCNQ$) method have been developed to tune the photo luminance performance of monolayer $MoS_2$[24].

## XPS characterization:

XPS spectra is sensitive to the chemical environment, and $MoS_2$'s crystal phase constitution can be determined by Mo 3d binding energy spectra[25]. As depicted in Figure 9(a-b), Hofmann *et. al.* study Mo 3d and S 2p spectra of 2H, 1T, amorphous $MoS_2$ changes before and after HER reaction, the XPS spectra show that Mo 3d and S 2p spectra of 2H $MoS_2$ unchanged, while Mo 3d and S 2p spectra shifted in 1T and amorphous $MoS_2$[26].

As depicted in Figure 9(c-f), McDowell *et. al.* made a detailed study on the interface of $MoS_2$ reaction with various metals (Li, Ge, Ag) on the interface by XPS characterization strategies, on the XPS study on $MoS_2$'s reaction under variant *in-situ* chemical deposit reaction[27], reaction between $MoS_2$ and Li, Ag, Ge were depicted. The XPS peak shift were attributed to interface reaction during metal deposition. To fitting Mo 3d spectra to multiple peaks corresponding to different phases, and get a brief phase constitution of 1T and 2H phase[28].

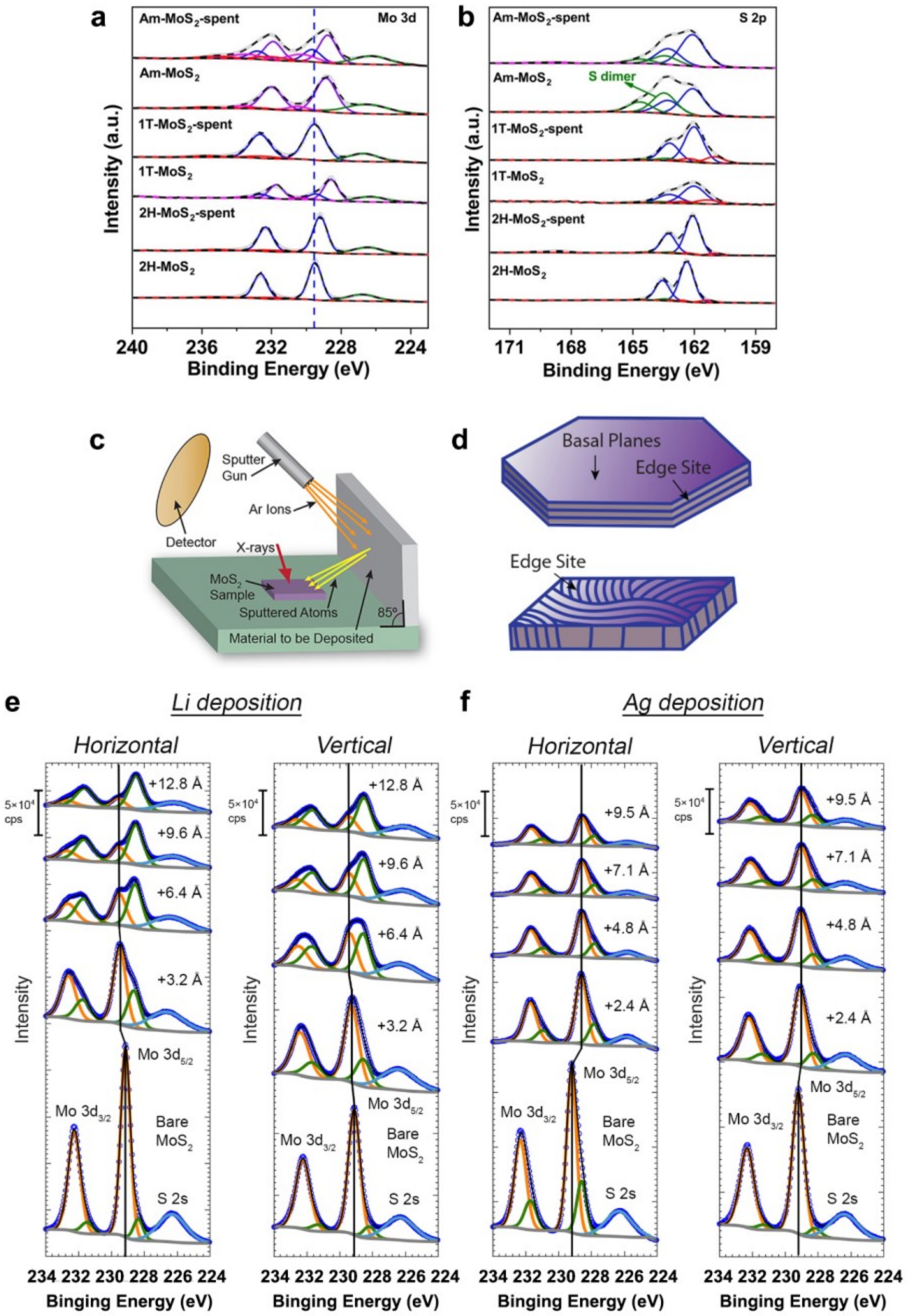

**Figure 9** (a) X-ray photoemission spectra of Mo 3d (a) and S 2p (b) before and after (spent) operando XAS measurements[26] (c) Schematic of a single $MoS_2$ crystal showing the position of basal planes and edge sites. (d) the layers are aligned along the electron beam as shown in the schematic below. (e) Evolution of Mo 3d peaks with progressive deposition of Li (e) and Ag (f)[27]. (Adapted with permission from ref. 26-27. Copyright 2019, Wiley-VCH Verlag GmbH & Co. KGaA, Weinheim)

## TEM characterization:

$MoS_2$ can be visualized to identify the crystal phase by TEM and attribute the TEM patterns to distinct crystal lattice. As depicted in Figure 10, in a tri-phase $MoS_2$ material, 1T, 2H, 3R lattice region in a single $MoS_2$ nanosheet can be identified by high resolute TEM[29]. In a typical literature, the Re doped single-layer $MoS_2$'s phase engineering between 2H and 1T were visualized[11]. What's more, grain boundaries and defects can be discerned in high resolute TEM. Appling to an environmental TEM, $MoS_2$'s phase conversion can be monitored. As depicted in Figure 11, J. Hong *et. al.* explored the atomic defects in $MoS_2$ monolayer, and found 5 types of antisite site by HADDF model TEM, successfully make an irrefutable proof on $MoS_2$'s defective structure[30]. And the electronic structures of different antisite defects were studied.

Distort 1T phase display zigzag metal chain attract much eyes for its unique properties: enhanced electrocatalytic performance, superconductivity, *etc*[31]. The structural relationship was clearly depicted by Zhu *et. al.*[32]

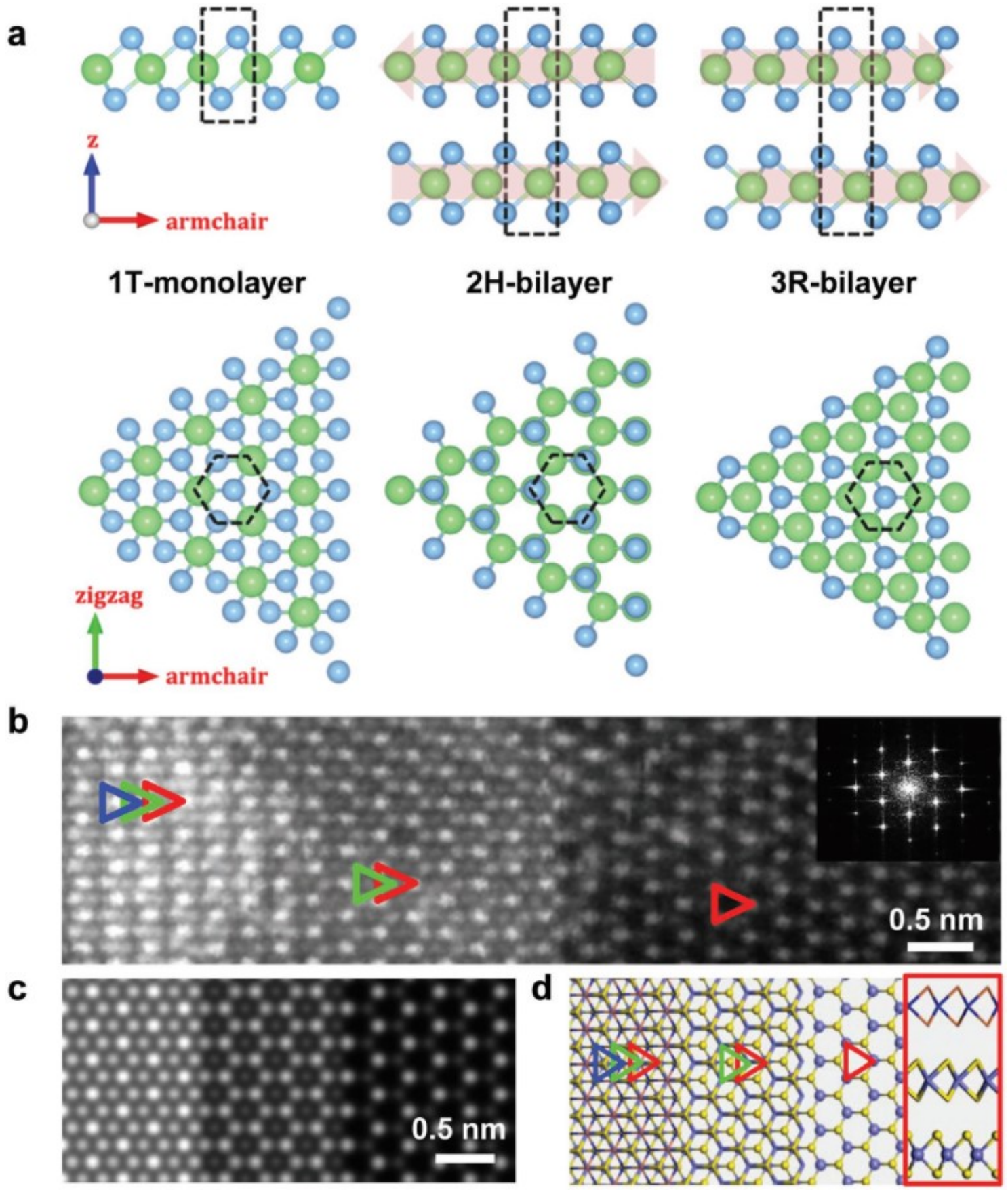

**Figure 10**. Schematic view of $MoS_2$ different phases and characterization of 3R phase $MoS_2$. (a) (Left to right): 1T, 2H phase (hexagonal symmetry), and 3R phase (rhombohedral symmetry) of $MoS_2$. (b) Atomic resolution STEM-ADF image of monolayer, bilayer, and trilayer 3R $MoS_2$ (from right to left) overlaid with corresponding diffraction pattern of 3R crystal. (c) Simulated image of STEM-ADF image of $MoS_2$. (d) Structure model corresponds to (b), which clearly shows the structure of different phase $MoS_2$. The picture in red frame shows the side view of monolayer MoS2[29]. (Adapted with permission from ref. 29. Copyright 2017, Wiley-VCH Verlag GmbH & Co. KGaA, Weinheim)

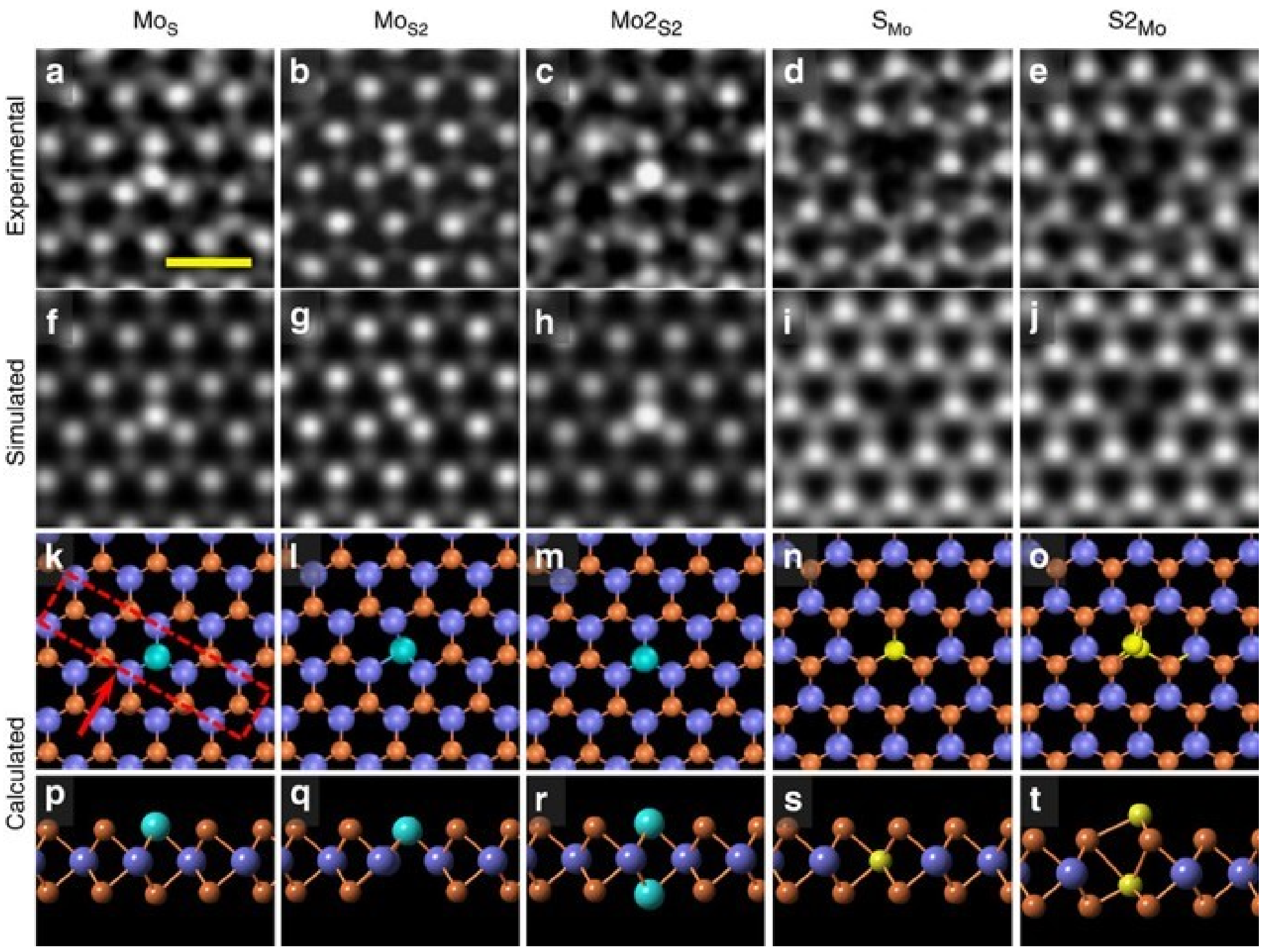

**Figure 11**. Different Mo antisite defect site[30]. (Adapted with permission from ref. 30. Copyright 2015, Nature Publishing Group)

**Raman spectra characterization:**

Raman spectra are widely applied to distinguish lattice structure[31], layer thickness of $MoS_2$[33]. 2H phase and 1T phase Raman spectra are distinct different and easily compared[31]. Hong Li *et. al.* established a detailed study on the $MoS_2$'s Raman spectra with different layer (1~4 layer), and Raman spectra were characterized with various laser lines (which are centered at 325, 488, 514.5, 532, 632.8 nm).[33] The relationship between the coupling of electronic transitions and the phonons were studied. The *operando* Raman spectra of amorphous $MoS_2$ were studied in detail by Yeo *et. al.*[34].

As depicted in Figure 12, tip-enhanced Raman spectroscopy (TERS) has been developed and successfully applied to make a closer focus on the localized $MoS_2$'s atomic structure, especially on the edge site structure which was identified as the catalytic site before[35].

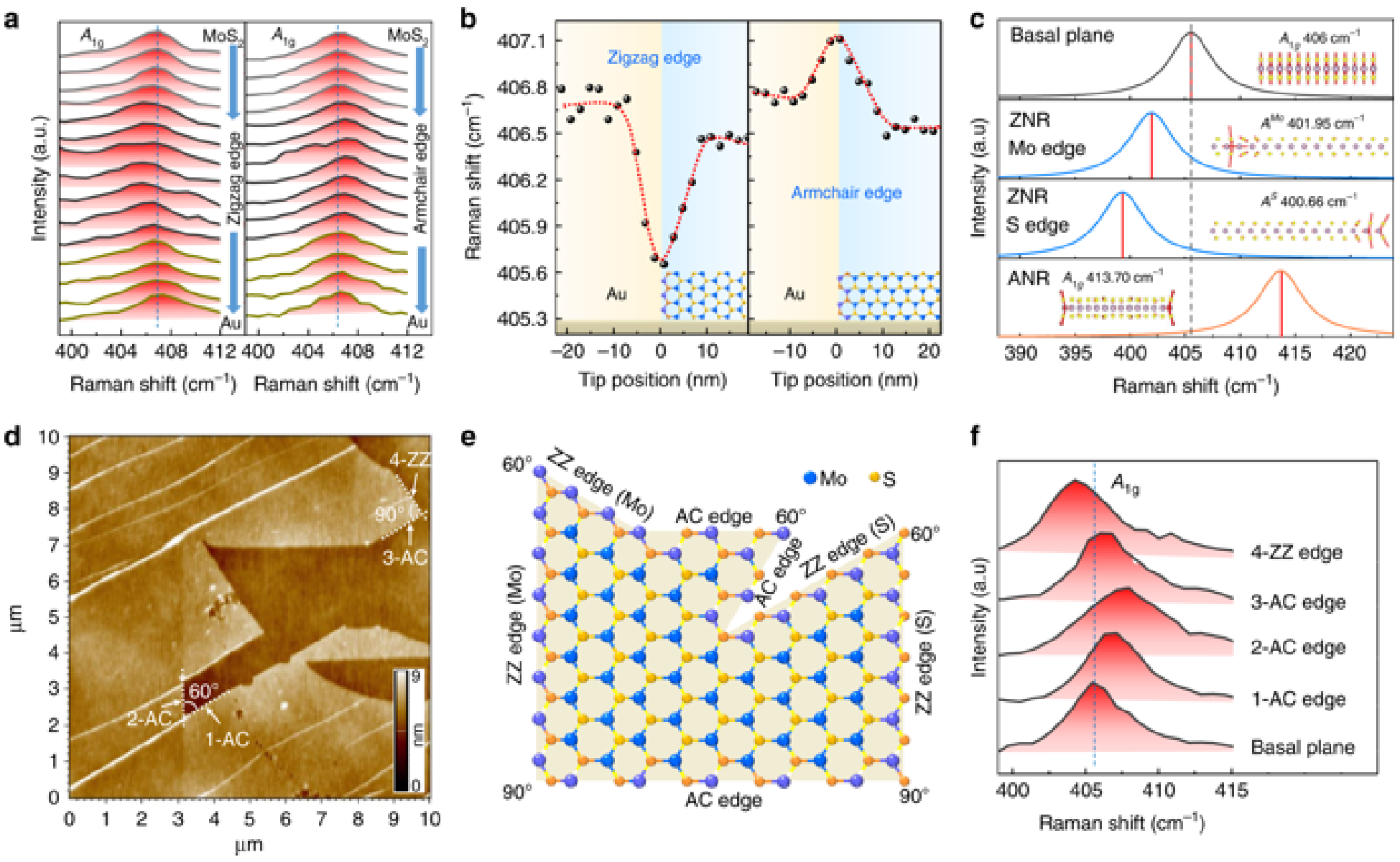

**Figure 12.** Effect of the edge structure on the peak position of the Raman $A_{1g}$ mode. (a) Typical line-trace TERS spectra of the zigzag edge (left panel) and armchair edge (right panel) in the spectral range of the $A_{1g}$ mode. (b) Plots of peak position with the tip position. (**c**) Calculated Raman spectra and lattice vibration of the basal plane, zigzag nanoribbon (ZNR, with a width of 3.59 nm) localized at the Mo and S edges, and armchair nanoribbon (ANR, with a width of 2.05 nm). (d) AFM image of a mechanically exfoliated 1 L $MoS_2$ with different edge angles on an Au substrate. (e) Illustration of the relationship between angles and edge structures of zigzag (ZZ) and armchair (AC) in 2 H $MoS_2$. (f) TERS spectra of four edges in the spectral range of the $A_{1g}$ mode marked in (d)[35]. (Adapted with permission from ref. 35. Copyright 2019, Nature Publishing Group)

X-ray absorption spectroscopy:

As depicted in Figure 13, the $MoS_2$ with 1T, 2H, as well as amorphous structures under HER were characterized, and the stability, bonding structure changes were clearly compared[26].

Zeleke *et. al.* applied XAS to study the interaction between $MoS_2$ and the carbonized polyacrylonitrile substrate, and find that low coordinating Mo catalytic site is responsible for higher catalytic performance[36].

*In-situ* X-ray absorption spectroscopy (XAS) has been widely applied in measurement of the nano catalysts' oxidation states, local environments during the real reaction conditions to footprint the changes in the catalysis process[37]. Hoffman *et. al.* make a detailed study on 1T, 2H, amorphous $MoS_2$ atomic bonding and electronic structures by utilizing Mo K edge XANES and Mo K-edge Fourier transform EXAFS (k2-weighted), the results clearly show that 1T and amorphous $MoS_2$ display short Mo-S bond than 2H $MoS_2$, which may afford for the better HER performance[37].

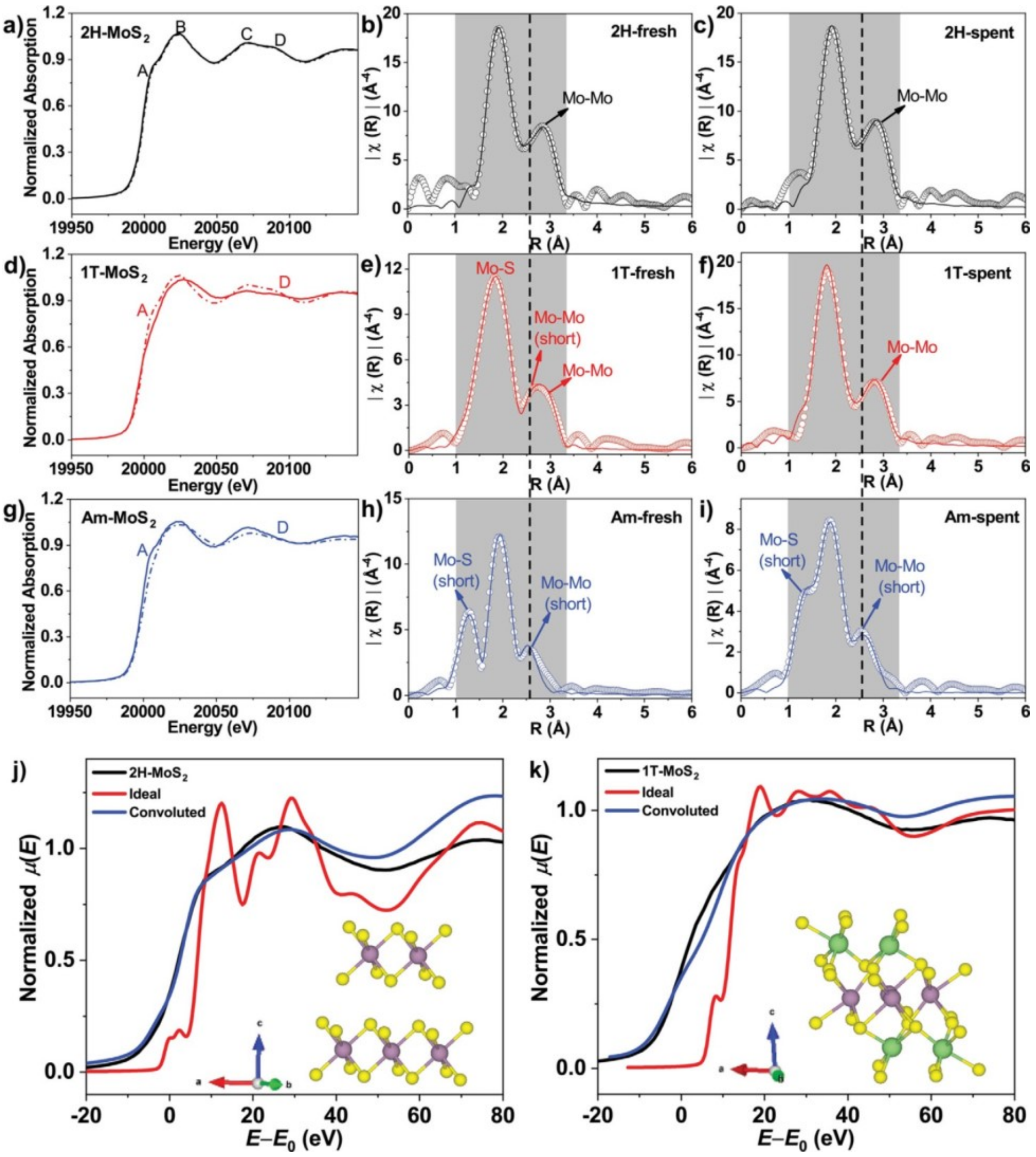

**Figure 13.** (a,d,g) Mo K-edge XANES spectra of 2H-$MoS_2$ (a), 1T-$MoS_2$ (d), and Am-$MoS_2$ (g) before (solid line) and after (dash line) stability test. b,e,h) Mo Kedge Fourier transform EXAFS (k3-weighted) of 2H-$MoS_2$ (b), 1T-$MoS_2$ (e), and Am-$MoS_2$ (h) before stability test. c,f,i) Mo K-edge Fourier transform EXAFS (k3-weighted) of 2H-$MoS_2$ (c), 1T-$MoS_2$ (f), and Am-$MoS_2$ (i) after stability test. j,k) Mo-K edge XANES spectra of experimental data (black curve) and calculated simulation based on hexagonal (j, inset) and monoclinic (k, inset) structure model; red curves represent simulated spectra whereas blue curves represent simulated spectra convoluted with the Mo 1s core-hole lifetime[26]. (Adapted with permission from ref. 26. Copyright 2019, Wiley-VCH Verlag GmbH & Co. KGaA, Weinheim)

**Strategies for $MoS_2$ engineering to elevate HER performance:**

Pristine $MoS_2$ displayed low catalytic performance for the low concentration of exposed catalytic site. Phase engineering, vacancy engineering, doping, basal plane activation, amorphous, heterojunction or boundary engineering are facile but remarkable strategies for promoting catalytic performance. To meet these demands, exposure edge catalytic sites, make active vacancies, elevate catalytic sites by activate basal plane, modification with cocatalyst were applied. And the relate performance enhancement were compared[38]. In some cases, the electrocatalyst achieve Pt-like performance. We make a brief comparation on different strategies on the performance.

**Basal plane engineering:**

$MoS_2$'s exposed low-coordinated edge sites are active catalytic site, since the exposed $MoS_2$ site exhibit moderate hydrogen adsorption behavior that facilitate the water splitting reaction[3]. As a comparation, pristine basal plane is inert for HER reaction.

To activate the basal plane, single atoms, heterojunctions were applied to introduce alien catalytic site. As depicted in Figure 14 (a-b), Re were introduced into $MoS_2$ to form alloyed $Re_{0.55}Mo_{0.45}S_2$ with distort 1T structure. The distort 1T phase was stabilized by Re, and the HER performance were elevated by activated basal plane[39]. Ruthenium doping and nanocarbon hybridization of $MoS_2$ have been synthesized by Xing Zhang *et. al.*, and the catalyst shows 50 mV at 10 mA/cm$^2$ (overpotential) with 62 mV/dec Tafel value[40].

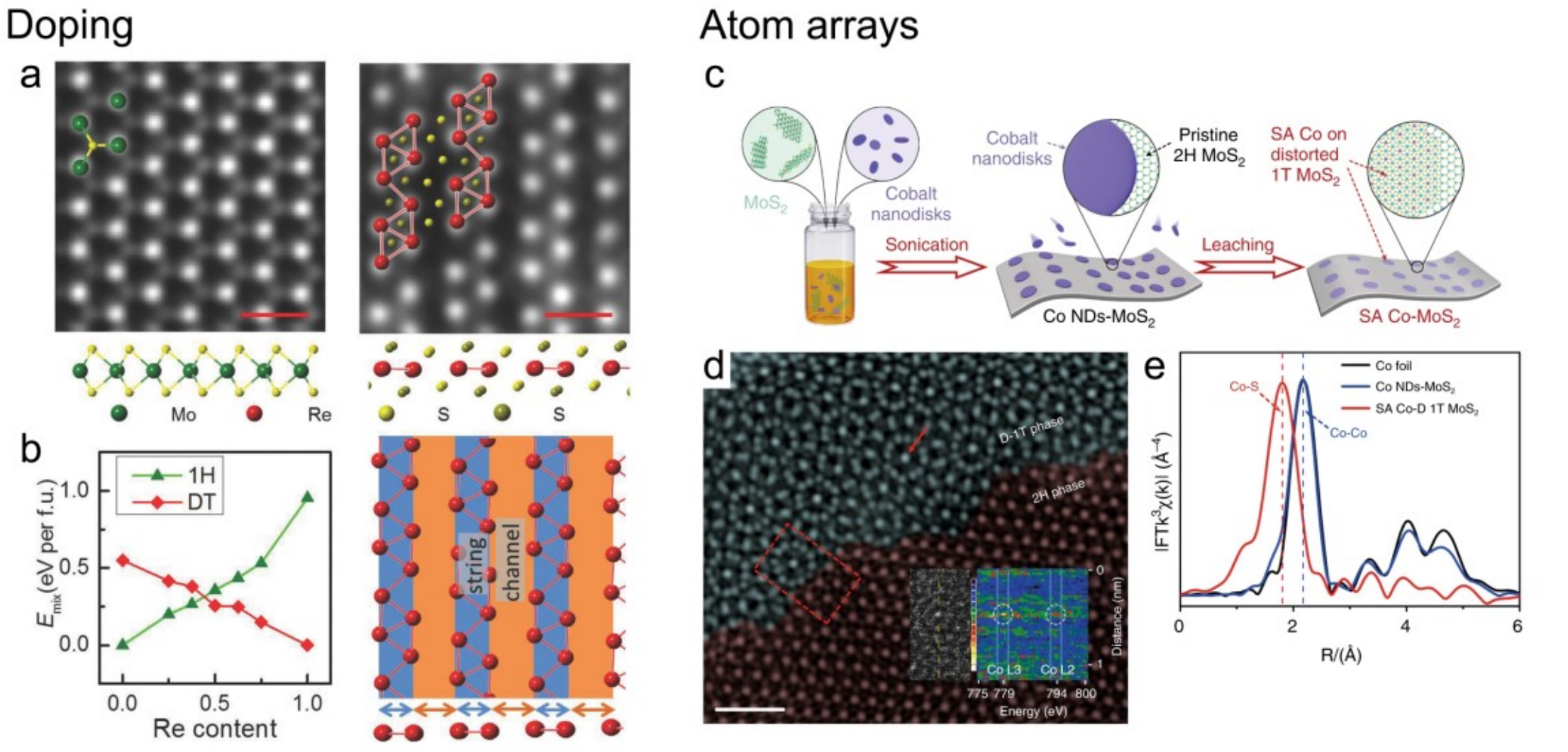

**Figure 14.** Basal plane activation by doping (a-b) and surface atom array modification (c-e) (a) Structural change diagram represent distort 1T phase induced by Re doping. (b) energy of materials phases relationship with variant Re doping concentration, and the diagram scheme of distort 1T phase with string and channel[39]. (Adapted with permission from ref. 39. Copyright 2018, Wiley-VCH Verlag GmbH & Co. KGaA, Weinheim) (c) Schematic diagram of the fabrication process for SA Co-D 1T $MoS_2$. (d) HAADF-STEM HRTEM spectra of 1T-2H heterojunction (e) FT-EXAFS spectra representation of strong interaction between metallic Co atoms and S atoms[41]. (Adapted with permission from ref. 41. Copyright 2019, Nature Publishing Group)

And by activating on the basal plane, Pt-like catalytic performance would be introduced[41]. As depicted in Figure 14 (c-e), cobalt single atom array modified basal plane were fabricated, since Co atoms on the basal plane act as extra catalytic active site, the catalyst exhibit elevated Pt-like HER performance[41]. The Co atom play as the catalytic site, which was identified by using $SCN^-$ to poison the Co site. When Co atom were coordinated with $SCN^-$ group, the catalysis performance was sharply reduced. Mo single-atom were loaded on monolayer $MoS_2$'s basal plane, and form unsaturated Mo atoms[42]. $Ni_2P$ modified on the basal plane of $MoS_2$ were synthesized by phosphidation of $NiMoS_4$ to afford $Ni_2P/MoS_2$. By coupled with conductive N-doped carbon support, Pt-like HER performance were achieved in acidic solution. The HER performance is superior to Pt/C catalyst at high current densities (>200 mA $cm^{-2}$)[43]. The author attributed the superior catalytic performance to $Ni_2P/MoS_2$ heterojunction, and the N doped carbon substrate (graphene or CNT) improve the stability and conductivity. $Au_{25}$ metal cluster were modified on $MoS_2$ to formulate the interface and display as cocatalyst in HER reaction[44], the onset potential arrived at -0.2 V vs RHE, the author attribute the enhanced catalytic performance comes from interfacial electronic interaction, and the thiolate (-SR), selenolate (-SePh) ligand on Au clusters show elevating effect to the catalytic performance.

Electrochemical method has been developed to activate the $MoS_2$[45]. Firstly, atomic layer deposition (ALD) were applied to coat $MoS_2$ with $TiO_2$, and form $TiO_2$ island on $MoS_2$'s basal plane, and $MoO_3$ were formed by interaction between $MoS_2$ and $TiO_2$. Then $TiO_2$ were leached by electrochemical activation method, and localized distort $MoS_2$ basal plane were produced and better HER performance were arrived. The P dopant dramatically reduced Mo valance charge and activate the inert basal plane. O assist P doping into 2H-$MoS_2$ has been developed and 130 mV onset potential and 49 mV $dec^{-1}$ Tafel slope were achieved[46].

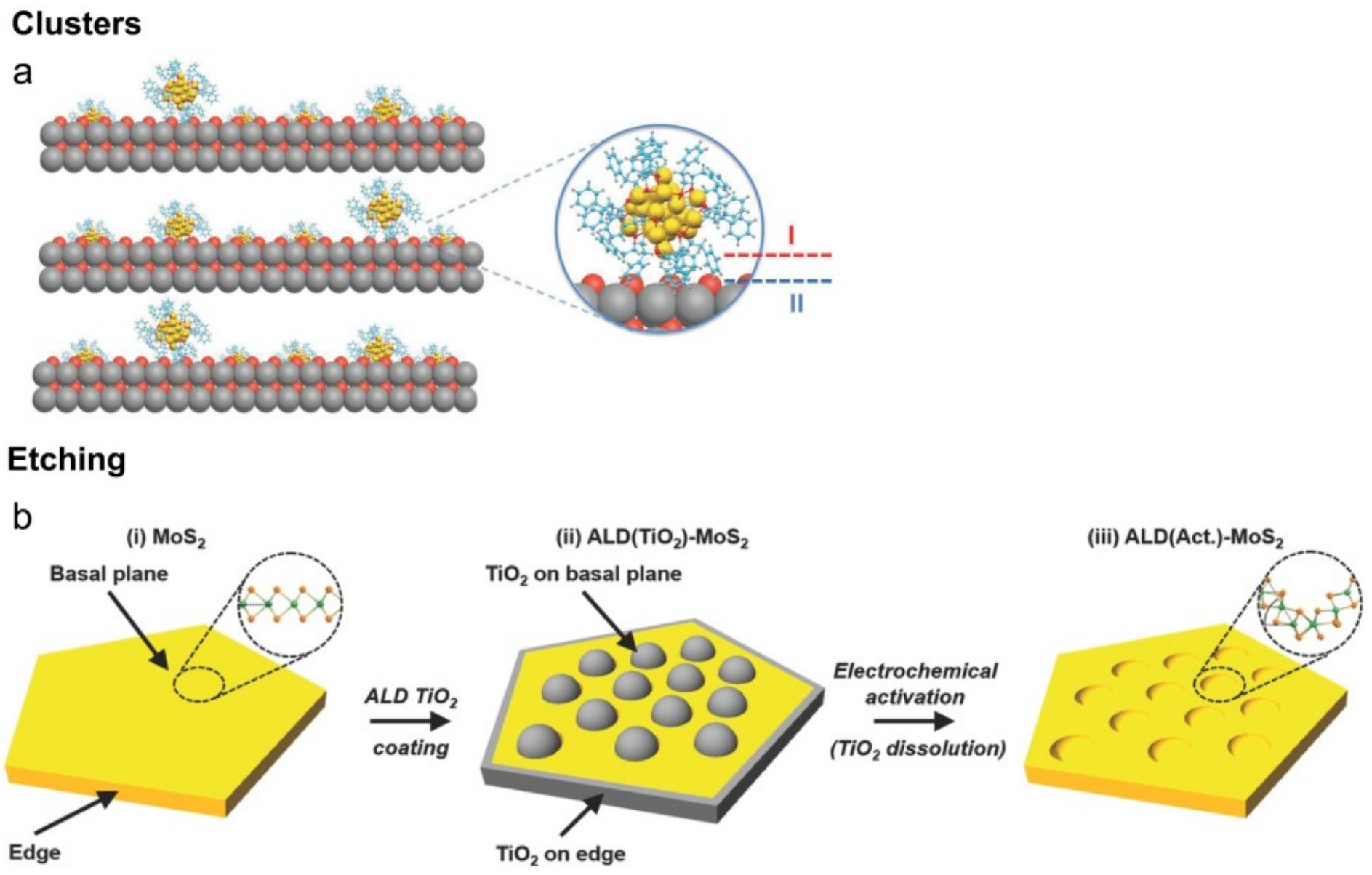

**Figure 15.** Basal plane cluster modification and etching (a) dual interfacial effect (Interface I and II) in $Au_{25}/MoS_2$ composite (yellow: Au; red/orange: S; blue: C; gray: Mo)[44]. (b) ALD $TiO_2$ coating on pristine $MoS_2$ and the enhancement of the hydrogen evolution reaction activity of ALD($TiO_2$) - $MoS_2$ catalysts via electrochemical activation[45]. (Adapted with permission from ref. 44-45. Copyright 2017, Wiley-VCH Verlag GmbH & Co. KGaA, Weinheim)

**Phase engineering by forming 1T phase $MoS_2$:**

$MoS_2$ compose three mainly different phase, which are 1T, 2H, 3R phases. 2H phase is mostly stable but exhibit with semiconducting performance, the limited conductivity is not friendly to electrocatalysis. It was also report that the 3R phase of $MoS_2$ depict better HER performance than 2H phase[17]. Whereas the metallic 1T phase shows the enhanced water splitting performance.

Hongli Zhu *et. al.* summaries the electrocatalysis performance of 1T metallic phase of $MoS_2$[32]. Treated with commercially available stable 2H $MoS_2$ with n-Butyllithium, large amount batch $MoS_2$ containing metallic 1T $MoS_2$ were arrived. However, arrived product by this method only get mixture of metallic 1T $MoS_2$ and semiconducting 2H $MoS_2$. Since 1T metallic phase were thermodynamic unstable, it will gradually transfer to stable semiconducting $MoS_2$ phase[47]. It shows that the extra electron injection into the $MoS_2$ layer are driving force for the phase conversion. The charge transfer kinetics and H adsorption on the active site were strongly escalated in 1T phase $MoS_2$ material.

A couple of strategies were applied to stabilize metallic 1T phase. Au-Pd-$MoS_2$ tri-composite heterointerface were fabricated to induce 1T phase $MoS_2$ by lattice mismatch effect[48]. Ir-$MoS_2$ heterointerface were fabricated, based on metal-support interaction by Ir adsorption, 2H phase were partly convert to 1T phase[49]. After performing a modulation calculation, the author discovered that when Ir atoms concentration was large enough, 1T phase were induced. And the author identified that Ir atom and the induced 1T phase were catalytic active site elevate the pristine performance. Wei Ding *et. al.* discovered that 1T phase $MoS_2$ and $WS_2$ can be synthesized under high magnetic field, and the resultant 1T-$MoS_2$ was stable for more than 1 year[50]. The magnetic field could transfer high energy to the material and adjust atomic and molecular alignment, and afford the expected morphology and phase. Gram batch scale synthesis for 1T phase $MoS_2$ has been developed by Li Song *et. al.*, the synthesis procedure was carried out by hydrothermal reaction between $(NH_4)_6Mo_7O_{24}$ and thiourea at 200 °C. When $NH_4^+$ were introduced into the layer space, the interlayer space was expended to 9 ~ 9.8 Å, and induce formation of distort 1T phase $MoS_2$ for interaction between $NH_4^+$ and $MoS_2$ layer[51].

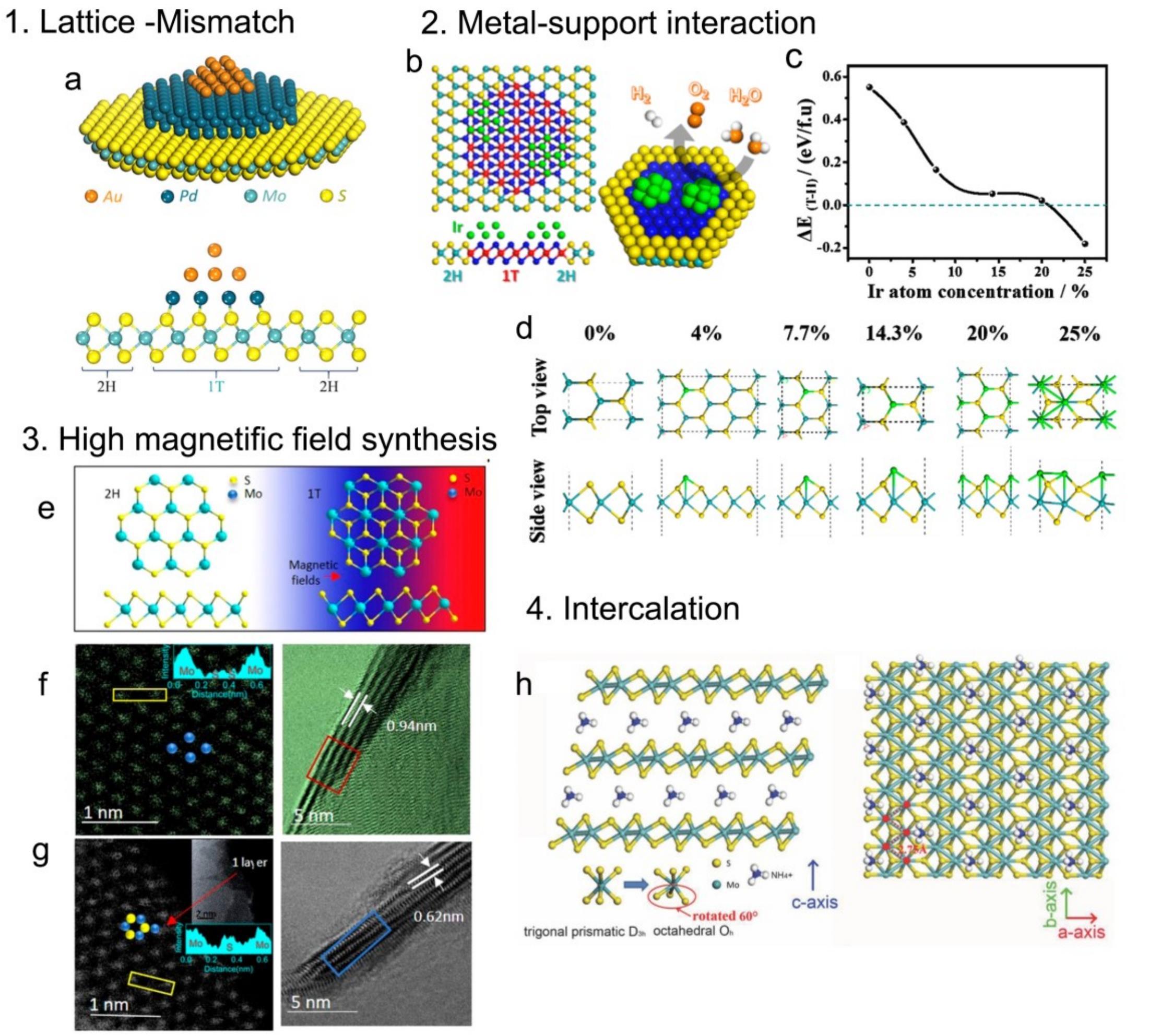

**Figure 16.** Strategies for fabricate phase convert $MoS_2$ (a) Lattice-mismatch interface induced phase conversion in Au/Pd-$MoS_2$[48] (Adapted with permission from ref 48. Copyright 2019, American Chemical Society.) (b-d) Metal support effect in Ir/$MoS_2$ induced phase conversion (b) Schematic diagram represents Ir cluster on the basal plane induced phase conversion (c) Energy diagram and (d) Theoretical model between Ir/1T-$MoS_2$ and Ir/2H-$MoS_2$ as a function of the atomic concentration of the adsorbed Ir[49]. (Adapted with permission from ref 49. Copyright 2019, American Chemical Society.) (e-g) hydrothermal synthesis of air stable 1T phase under high magnetic field [50]. (Adapted with permission from ref 50. Copyright 2019, American Chemical Society.) (h) interlayer species interaction induced phase conversion. schematic diagram representing of $NH_4^+$ intercalated $MoS_2$ induced distort 1T phase and the corresponding zig-zag chain structure[51]. (Adapted with permission from ref. 51. Copyright 2015, Wiley-VCH Verlag GmbH & Co. KGaA, Weinheim)

**Doping effect:**

Doping atoms can induce phase conversion[11]. In a Re-$MoS_2$ material, Re atom substitute $MoS_2$ shows that phase conversion are much more facile to achieve. Various metal or non-metal atoms were solely or co-introduced into $MoS_2$ lattice, the introduced atom can modify the electronic structure of $MoS_2$ and adjusted the physical properties, induce vacancies, and spontaneously create new catalytic site. In some cases, metal/non-metal co doping introduce superior Pt-like HER performance under high current density. Doping engineering strategies have shown modulate and adjust the physical structure as well as electronic structure of $MoS_2$, strongly elevate the HER performance.

Metal atoms (Pt, Co, Zn, Ni, W *etc.*) doping effect

Various metal atoms were doped into $MoS_2$ lattice to elevate the HER performance. According to the theoretical calculation that the dopant can adjust the H adsorption energy to a moderate level and facilitate HER performance[52-53]. It has been found out that the substitute doping of transition metal ions like Re, Tc, Mn stabilized 1T metallic phase for the electron donating effect induced more stabilized 1T electronic structure[54].

Zn doped $MoS_2$ was synthesized and the HER performance show that the Zn dopant exhibit superior effect on accelerate the HER reaction (0.13 V overpotential)[55]. The authors attributed the large promotion effect to energy level matching effect as well as morphological effect via thermodynamic and kinetic route. Fe doped $MoS_2$ were fabricated on Ni foam by simply one-pot solvothermal reaction, and the resultant catalyst display HER performance with a 173 mV overpotential in 0.5 M $H_2SO_4$ (10 mA $cm^{-2}$), 153 mV in 1 M KOH (10 mA $cm^{-2}$), and also show OER performance with a overpotential at 230 mV (1 M KOH, 20 mA $cm^{-2}$)[56]. Co doped $MoS_2$ nanosheets bonded on carbon substrate were synthesized and the hybrid catalyst display elevated HER performance[57]. Bao *et. al.* report the single Pt doped few-layer $MoS_2$, whereas Pt is not working as the catalytic site[53].

High valance W metal ions were incorporated into $MoS_2$ and hybrid with $MoO_2$ and CNT, the W-$MoS_2$/$MoO_2$/CNT exhibit a small Tafel slope of 44 mV $dec^{-1}$ close to 20 % Pt/C (34 mV $dec^{-1}$)[58].

Non-metal atoms (N, P, O, C, Se) doping effect

Non-metal doping like C, O, P, Se were doped into $MoS_2$, and introduce superior HER performance. N doped $MoS_2$ were successfully synthesized by Xiao *et. al.*, in their work N atoms were doped on the edge as well as on basal plane, spontaneously boost the catalytic HER performance and exhibit 121 mV (overpotential, 100 mA $cm^{-2}$) and 41 mV $dec^{-1}$ Tafel slope[59] (Figure 17). N dopant, $PO_4^{3-}$ intercalated $MoS_2$ were synthesized by Deng *et. al.*[25], the N and $PO_4^{3-}$ induced phase conversion to 1T-$MoS_2$ (ca. 41 % 2H-$MoS_2$ were convert), which was higher than that of N doping (ca. 28 %) or $PO_4^{3-}$ intercalation (ca. 10 %). When combined with graphene, the hybrid catalyst emerged 85 mV (overpotential, at 10 mA $cm^{-2}$) and 42 mV $dec^{-1}$ (Tafel slope).

As depicted in Figure 17, O doped $MoS_2$ nanosheet were synthesized by hydrothermal reaction between $(NH_4)_6Mo_7O_{24}$ and extra thiourea. The doped oxygen atom help expend the interlayer space to 9.5 Å, confirmed by XRD as well as HRTEM results[60]. Tapasztó *et. al.* find out that by exposure of $MoS_2$ monolayer under ambient condition, $MoS_2$ suffers from a kinetically slow oxygen-substitution reaction to introduce O atoms on the basal plane and can get arrived $MoS_{2-x}O_x$[61].

Se were doped into the lattice of $MoS_2$ to form $MoS_{2(1-x)}Se_{2x}$ with tunable concentration of Se atoms without phase conversion[62] (Figure 17). The catalyst displayed strong stability without negligible activity loss up to 10000 cycles and the overpotential is small in the range of 80~100 mV.

As depicted in Figure 17, P doped $MoS_2$ were synthesized by annealing bulk $MoS_2$ with red phosphorus at 750 °C to form $MoS_{2(1-x)}P_x$ Solid Solution materials[63]. The best HER performance can be arrived at the constitution of $MoS_{0.94}P_{0.53}$ with an overpotential of 150 mV to reach the current density of 10 mV/cm$^2$. P doped $MoS_2$ (with interlayer distance of ~9.0 Å) were also synthesized by Liu *et. al.* and the HER performance exhibit ~43 mV (overpotential, 10 mA/cm$^2$) and 34 mV/dec (Tafel slope) as well as long-term catalytic stability[64].

Zang *et. al.* synthesized C doped $MoS_2$ by controlled sulfurization of $Mo_2C$, and the prepared $MoS_2$ displayed an unprecedented overpotential of 45 mV at 10 mV/cm$^2$ in the alkaline environment, which is just a bit lower than commercial Pt/C[65] (Figure 17). The electronical as well as coordination structure were significantly changed by corporation of carbon atoms, the authors attribute the elevated HER performance to the orbital modulation, that C induced empty 2p orbitals perpendicular to $MoS_2$'s basal plane, facilitate water adsorption and dissociation.

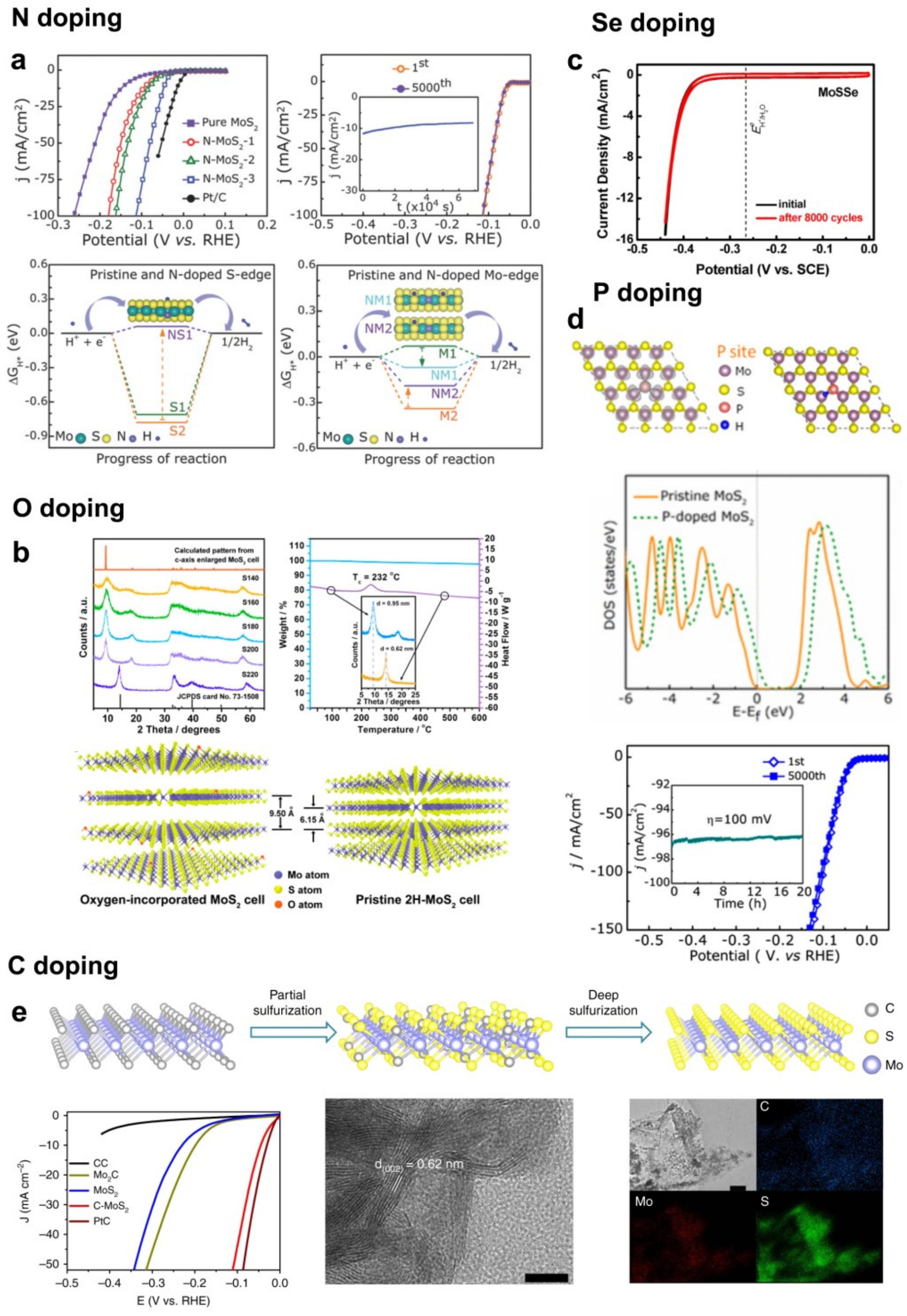

**Figure 17.** Non-metal atoms (N, P, O, C, Se) doping effect of $MoS_2$ towards elevated HER performance. (a) N doping. The schematic illustration of the synthesis of C–$MoS_2$ and $MoS_2$[59]. (Adapted with permission from ref. 59. Copyright 2017, Wiley-VCH Verlag GmbH & Co. KGaA, Weinheim) (b) O doping. The XRD patterns of O incorporated $MoS_2$ and the structure diagram scheme.[60] (Adapted with permission from ref 60. Copyright 2019, American Chemical Society.) (c) Se doping. CV curves of MoSSe before and after 8000 potential cycles[62] (Adapted with permission from ref. 62. Copyright 2015, American Chemical Society.) (d) P doping. CV curves of P doped $MoS_2$ before and after 5000 potential cycles[64] (Adapted with permission from ref. 64. Copyright 2017, American Chemical Society.) (e) C doping. The schematic diagram of synthesis of C doped MoC[65]. (Adapted with permission from ref. 65. Copyright 2019, Nature Publishing Group)

**Co-doping effect:**

Co-doping method exhibit dual activation ability by activation on different catalytic site. Co-doping strategies were employed by D. Deng *et. al.*, doping with Se on the basal plane surface and Co in the inner layer, the HER performance exhibit with a overpotential of 132 mV at 10 mA $cm^{-2}$[66]. The author consider that the inner-layer Co-doping and the surface Se-doping exhibit a synergic effect to improve the HER performance, attributing to the enriched catalytic site and hydrogen adsorption optimizing. And in current density at 1000 mA $cm^{-2}$, the Co, Se co-doped $MoS_2$ overpotential reach 382 mV, much lower than commercial 40 % Pt/C, and working continuously for at least 360 h without any decay. Huang *et. al.* found that Ni, O co-doped 1T phase $MoS_2$ catalyst exhibit Pt-like hydrogen evolution performance in alkaline solution[67]. In which Ni, O were synergistic doped in 1T $MoS_2$, and the overpotential at lower current density (< 80 mA $cm^{-2}$) exhibit better performance than commercial 20 % Pt/C. Xiong *et. al.* discovered cobalt doped $MoS_2$ show good HER performance under acidic as well alkaline solution[68].

Mn, N co-doped $MoS_2$ was synthesized by Su *et. al.*, the electronic structure of the doped $MoS_2$ material exhibit strongly enhanced HER performance with the overpotentials of 66 and 70 mV at 10 mA $cm^{-2}$ in alkaline and phosphate-buffered saline media, respectively[69]. O, P co-doped $MoS_2$ was synthesized by Dongdong Zhu *et. al.* via hydrothermal treating of $MoS_2$ with $NaH_2PO_2$, and the experimental results show that the co-doped $MoS_2$ display better HER performance than O doped $MoS_2$[70]. P, Se co-doped $MoS_2$ were fabricated using $C_4H_{14}N_3PS$ and $Na_2SeO_3$ as additive, the resultant P, Se co-doped $MoS_2$ display better performance than mono-doped $MoS_2$[71]. F, N co-doped $MoS_2$ were synthesized to activate basal plane and onset overpotential of 110 mV and 57 mV $dec^{-1}$ Tafel slope were achieved[72].

**Vacancy engineering:**

Vacancies are typical catalytic site that not limited to HER reaction[73]. Mo vacancies and S vacancies were fabricated in $MoS_2$ to elevate the catalytic HER performance. Sulfur vacancy can be introduced by electrochemical reduction[74], etch by $H_2$[75], $H_2O$ vapor[76].

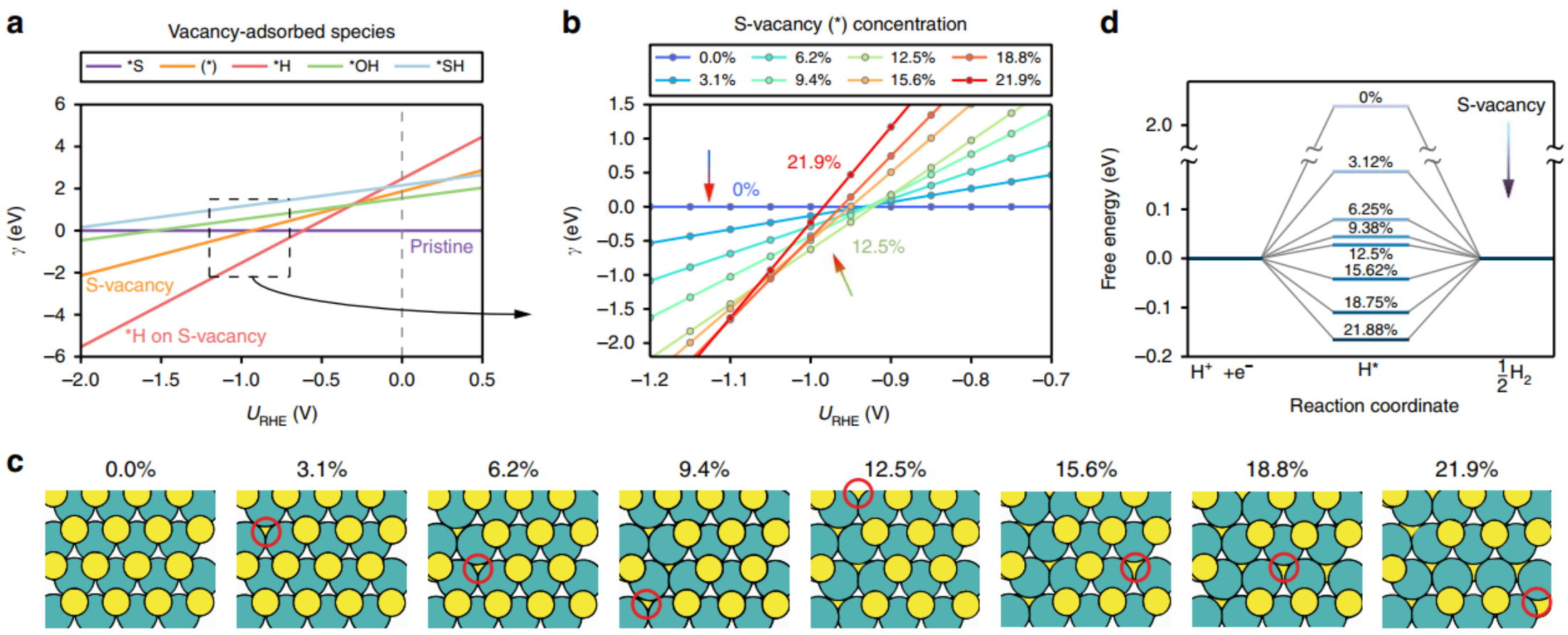

**Figure 18.** (a) Surface energy per unit cell for 2H-$MoS_2$ as a function of applied potential for the basal plane of 2H-$MoS_2$ with different adsorbate species at a fixed sulfur vacancy (3.1%). (b) Surface energy per unit cell for a range of S-vacancy concentrations, without any adsorbates. The concentration of S-vacancies varies from 0 to 21.9% within the narrow range of 1.0 V to 1.1 V. (c) When the S-vacancy sites are generated in succession, S-vacancies are most stable when formed next to an existing S-vacancy. (d) Free energy diagram for the HER on S-vacancy sites[74]. (Adapted with permission from ref. 74. Copyright 2017, Nature Publishing Group)

As depicted in Figure 18, Charlie Tsai *et. al.* applied the electrochemical strategies to introduce sulfur vacancies into $MoS_2$ materials, and the HER performance were examined[74]. By applying different potential, the extent of desulfurization was varied, and HER performance were spontaneous alternated.

Sulfur vacancy were fabricated by William A. Goddard III *et. al.*, they made a detailed study on the HER reaction on the sulfur vacancies. They discovered that the sulfur vacancies are favored to elevate the HER performance for that the transition state energy is closer to the product, and it can get a more moderate Tafel slope[77]. Cao *et. al.* synthesized the $MoS_2$ materials and introduced variant amount of sulfur vacancies by simply calcinate molybdenum chloride ($MoCl_5$) with sulfur under 850 °C, they find that the 7~10 % sulfur vacancies density and better crystallinity help entrance towards better HER performance[78].

As depicted in Figure 19, Xin Wang *et. al.* introduced homogeneously distributed sulfur vacancies by $H_2O_2$ etching method, by systematically adjusting of etching time, temperature, and $H_2O_2$ concentration, various S-vacancy state can be achieved[79]. This systemic study shows a better understanding between the structure, concentration of sulfur vacancy and the relate HER performance. D. Voiry *et.al.* made a systematic study on the variant amount sulfur vacancies relation with the HER performance. They clarified that when point S vacancy were introduced to $MoS_2$ rapid elevating of HER performance can be tracked, while more S atoms were removed (atom ratio: $S/Mo > 1.7$), the catalyst is dominated by the uncoordinated Mo site and S vacant[80]. Sulfur vacancies concentration were introduced by regulating reduction conditions (atmosphere and temperature).

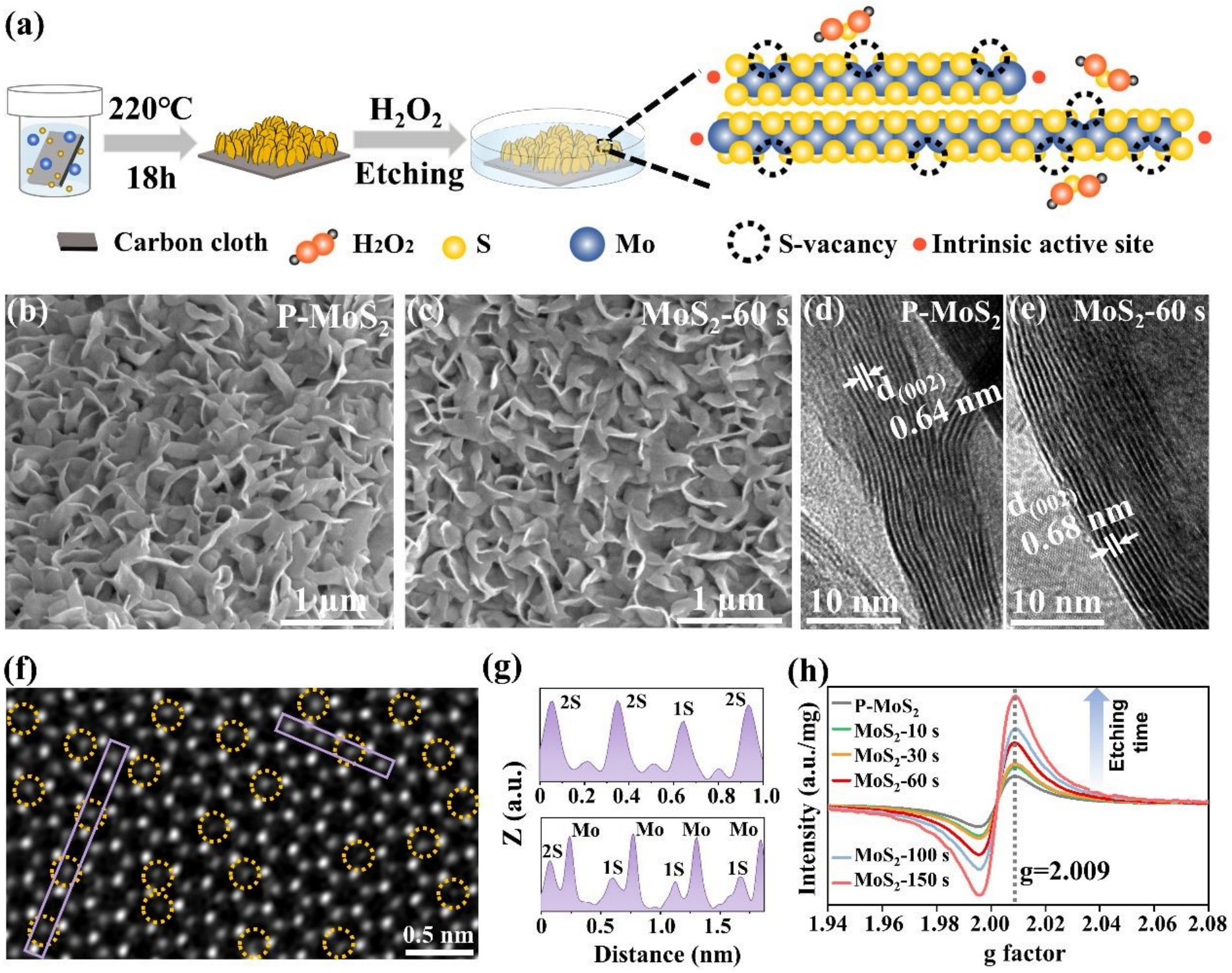

**Figure 19.** (a) Schematic of the chemical etching process to introduce single S-vacancies. b-e) SEM and HRTEM images of (b, d) P-$MoS_2$ and (c, e) $MoS_2$. f–g) The STEM image together with the line profiles extracted from the areas marked with purple rectangles of a CVD-grown monolayer $MoS_2$ flake film after etching. The yellow dotted circles represent the S-vacancies. h) EPR spectra of etched $MoS_2$ with different etching durations[79]. (Adapted with permission from ref. 79. Copyright 2020, American Chemical Society.)

As depicted in Figure 20, pristine 2H $MoS_2$ were annealed in Zn vapor to forming sulfur vacancies and Zn atom dopant. The results shows that the Zn can induce sulfur vacancies and Zn dopant can also active the adjacent S vacancy site, and elevate HER performance[81]. By reduction reaction with Zn, $MoS_2$ were cracked into small pieces (pristine 100~200 nm sheet were cracked into sub-25 nm sheet).

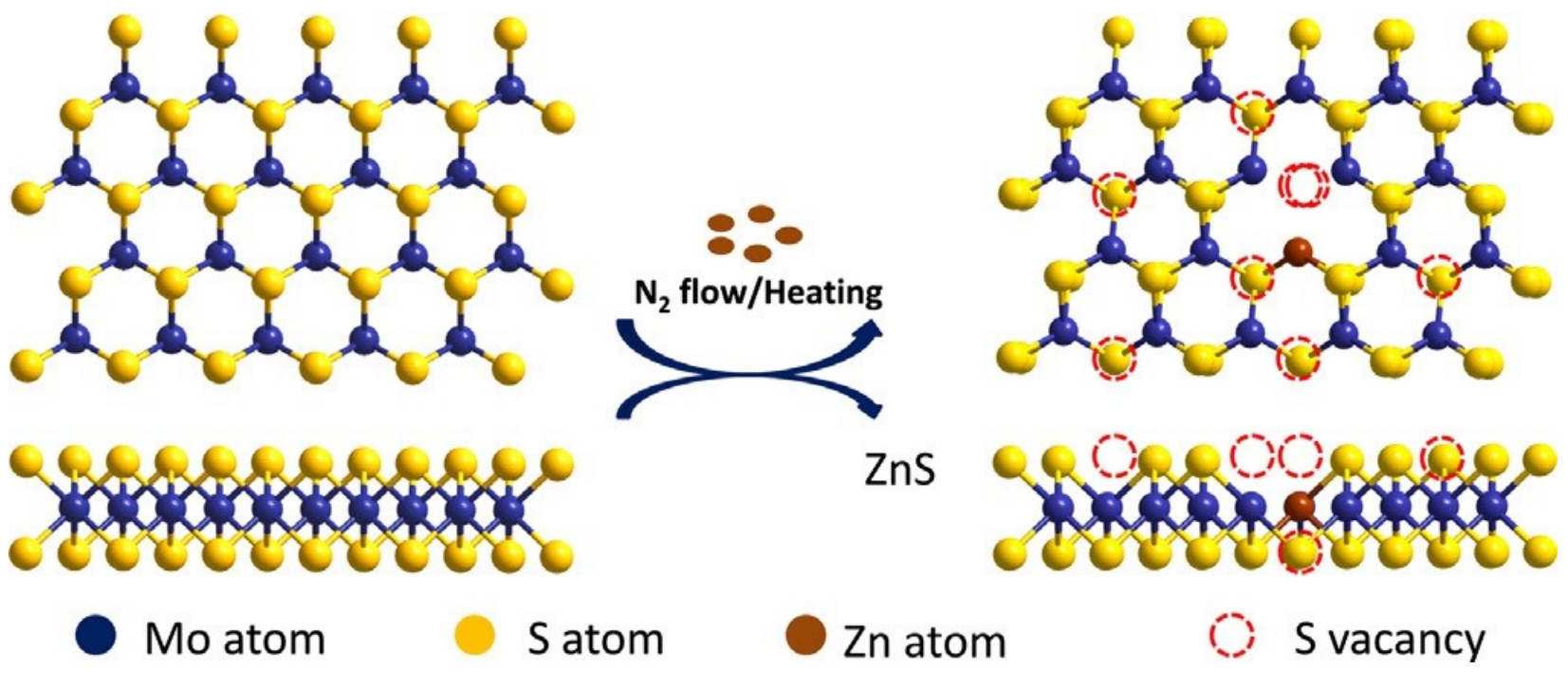

**Figure 20.** Diagram of the experimental creation of S vacancies in a 2H-$MoS_2$ nanosheet by Zn vapor annealing[81] (Adapted with permission from ref. 81. Copyright 2019, Wiley-VCH Verlag GmbH & Co. KGaA, Weinheim).

Strained sulphur vacancies were introduced into $MoS_2$ material and HER performance were discovered, it was found that the performance was elevated by carried out the HER reaction under different sulphur concentration (0~21.9 %), and the best catalytic performance is not followed with higher concentration of vacancies. The theoretical results show that the hydrogen adsorption energy can be modified by adjust the strain on S vacancy, and elevate HER catalytic reaction[82].

16 types of different structural defects including point defects, grain boundaries were analyzed by first-principle calculations to clarify how the intrinsic vacancies influence the $MoS_2$'s HER performance[83]. And using amendatory band-center model to examine different HER performance along various kinds of vacancies.

**Interlayer space expending and intercalation:**

Edge-site low coordinated atoms were regard as the best catalytic site, since interlayer engineering can induce the local chemical environment and electronic structure changes, the relate catalytic performance were modulated.

Intercalation of $Li^+$ into $MoS_2$ layer space is widely applied for $MoS_2$ exfoliation. What's more, chemical intercalation is a traditional and welcomed strategy for metallic $MoS_2$. G Eda. *et. al.* used *n*-Butyllithium compound to exfoliate $MoS_2$ in hexane under Ar atmosphere[22]. Polymers, molecule, cationic species, organometallic species has been intercalated into the layer of $MoS_2$ to expend the layer distance[84-86].

The expend layer can reduce the resistance during the ion in the layer space. By intercalate by poly(ethylene oxide), the interlayer distance is expanded to 1.45 nm[87]. Hydrothermal reaction using $(NH_4)_6Mo_7O_{24}\cdot 4H_2O$ with thiourea as reactant would yield $MoS_2$ with a 0.95 nm interlayer space[60]. The relative study shown that $NH_4^+$ were intercalated into the layer space[51].

Increasing interlayer distance would impact the band energy of $MoS_2$ and electronic structure. As depicted in Figure 21, Sun *et. al.* synthesized the enlarged $MoS_2$ with an interlayer distance of 9.4 Å, in this way the stacked $MoS_2$ with enlarged interlayer's perform like the freestanding monolayer $MoS_2$. Theoretical results shown that the expended layer distance induce the band gap shift upwards along the Fermi Level by ~0.1 eV (Figure 21b), and enlarged layer distance facilitate the H adsorption on $MoS_2$'s edge site[38, 86].

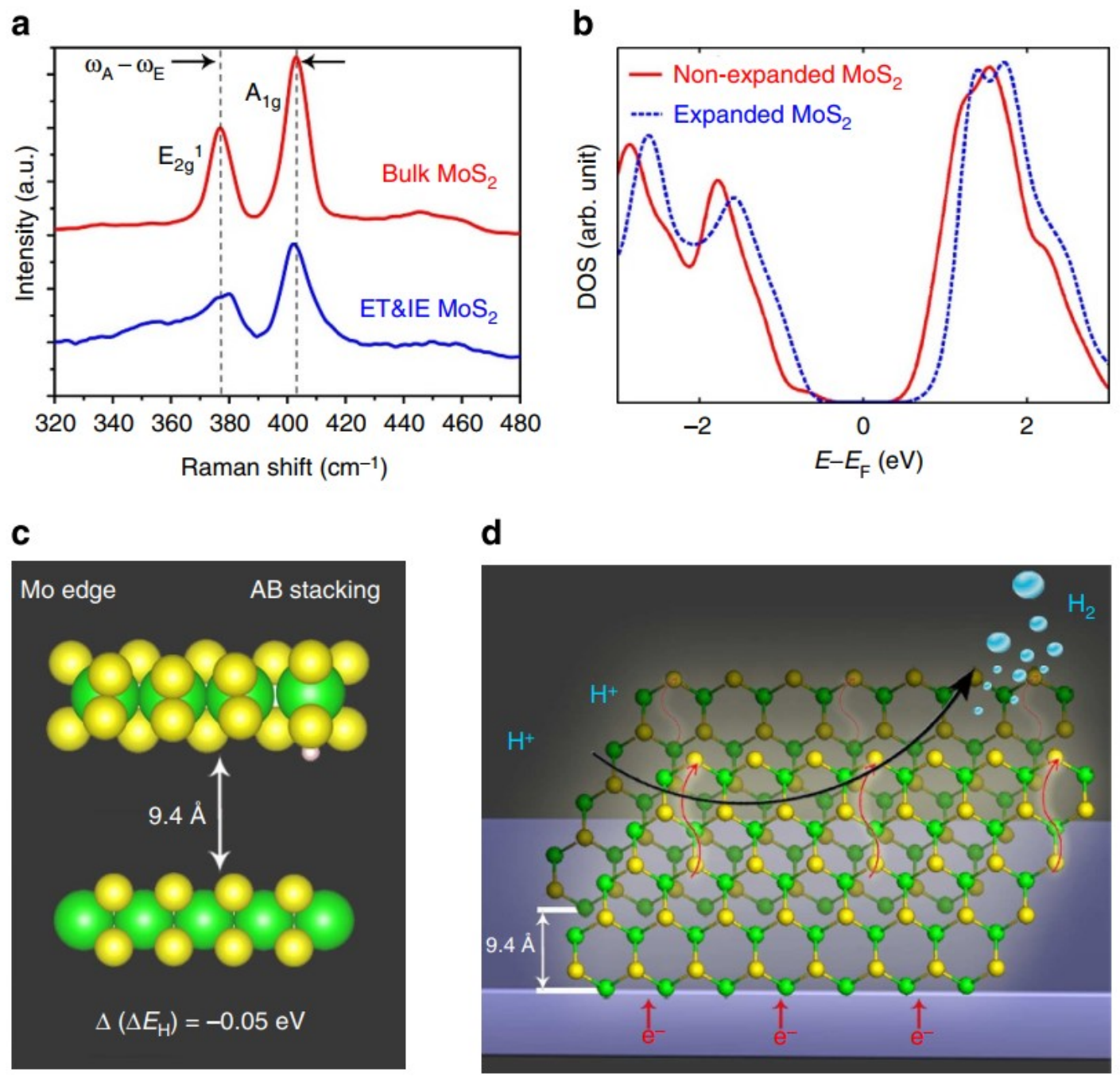

**Figure 21.** (a) Raman spectra (b) The projected DOS of bulk $MoS_2$ and ET&IE (edge-terminated and interlayer-expanded) $MoS_2$. (c) DFT calculation of the change in hydrogen adsorption energy. (d) Schematic representation of the edge-terminated $MoS_2$ on glassy carbon electrode for HER[38]. (Adapted with permission from ref. 38. Copyright 2015, Nature Publishing Group)

Ultrathin $MoS_2$ with a 9.5 Å interlayer distance coupled to N-doped reduced graphene oxide was synthesized by a one-step hydrothermal reaction[88]. As depicted in Figure 22, the 9.5 Å larger interlayer present a preferable $\Delta G_{H*}$ value of -0.052 eV, while the pristine $MoS_2$ with 6.3 Å shows -0.201 eV. The enlarged layer distance display better proton/electron adsorption and hydrogen release step, and superior HER performance was verified.

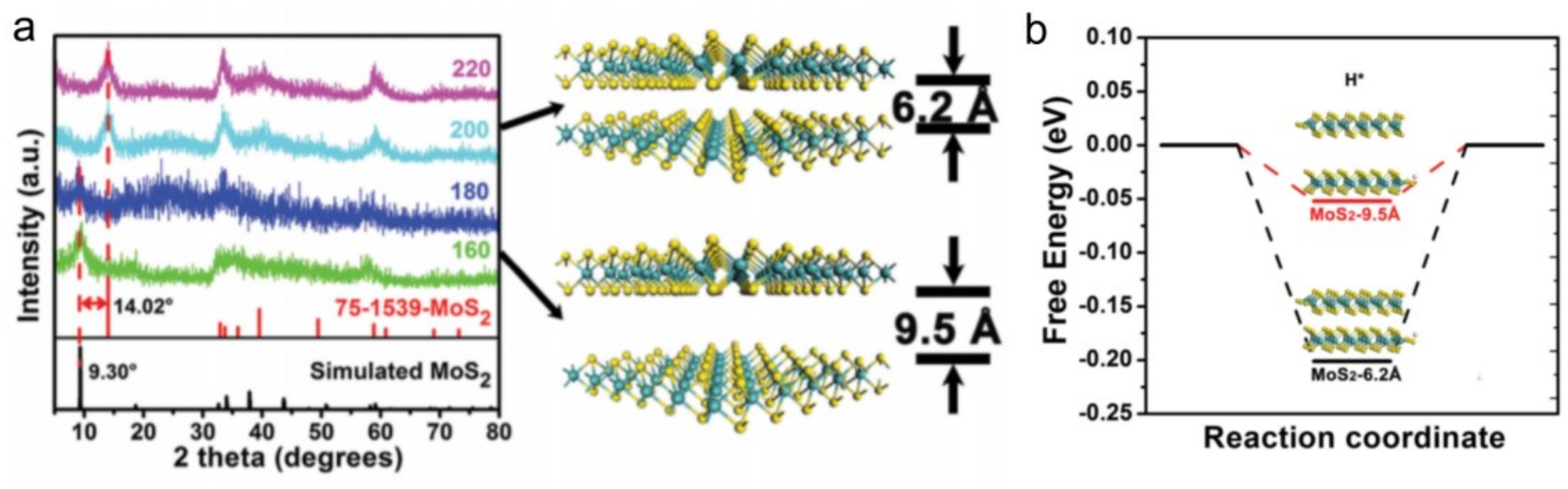

**Figure 22.** (a) XRD patterns of $MoS_2$/N-RGO-T nanocomposite prepared at different temperatures from 160 to 220 °C. At the right, structural models of $MoS_2$ with the interlayer spacing of 6.2 and 9.5 Å, respectively. (b) Calculated free energy diagram for HER on $MoS_2$-6.2 Å and $MoS_2$-9.5 Å[88]. (Adapted with permission from ref. 88. Copyright 2016, Wiley-VCH Verlag GmbH & Co. KGaA, Weinheim)

**Edge exposure engineering:**

Edge sites are catalytic active site for the HER reaction, exposure of large amount of edge site are fundamental strategies for achieve high-performance HER catalysts. Self-assembly, novel synthesis method like chemical vapor deposition (CVD)[89-90], surfactant-assist low-temperature solution process, hierarchical nanostructure fabrication strategies were applied to exposure with more active edge site. As depicted in Figure 23, Cui *et. al.* deposit 5 nm Mo layer on quartz, $SiO_2$/Si substrate and then Mo react with S vapor to afford vertically aligned $MoS_2$[90]. Decreasing of the Tafel slope to 49 % can be achieved by selective steam etching method by expose with more active edge site[76, 91-92]. As depicted in Figure 24, by adjusting precursor composition, Ruitao Lv et. al. using electrospinning strategies fabricate carbon nanofibers with large amount of exposed $MoS_2$ edge site[91].

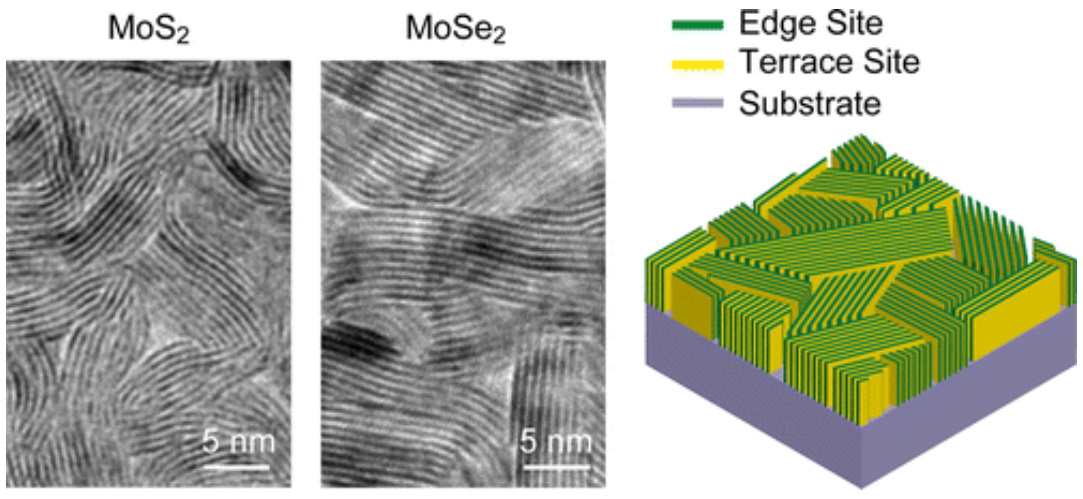

**Figure 23.** (left) TEM image of a $MoS_2$, $MoSe_2$ films with edge-terminated structures. (right) Idealized structure of edge-terminated molybdenum chalcogenide films with the layers aligned perpendicular to the substrate[90] (Adapted with permission from ref. 90. Copyright 2013, American Chemical Society.)

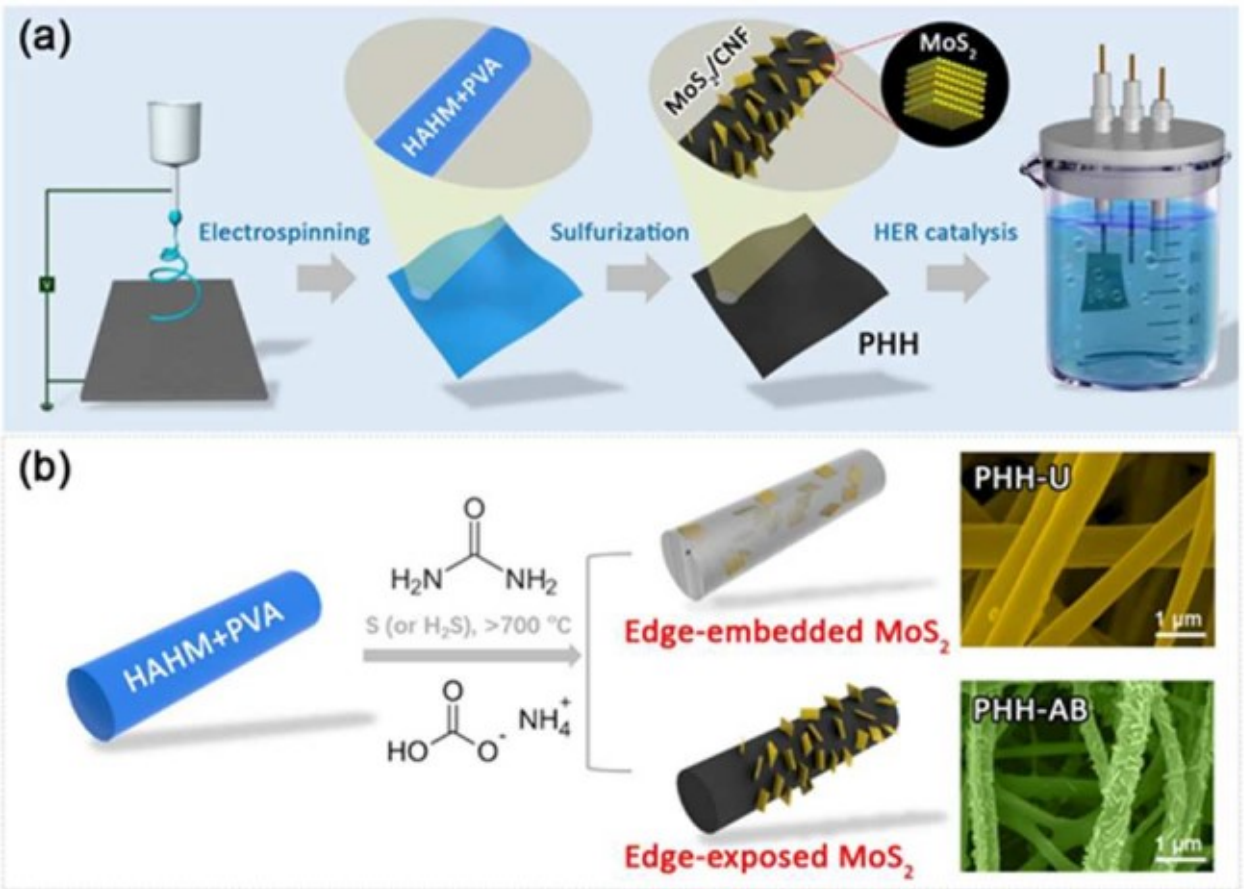

**Figure 24.** (a) Schematic illustration of electrospin synthesis of $MoS_2$/CNF nanocomposite and HER catalysis. (b) Controlled growth of $MoS_2$ nanoplates in the CNFs from different precursors, and the corresponding structures with $MoS_2$ nanoplates embedded inside (the top right SEM image) or exposed outside (the bottom right SEM image) of CNFs[91]. (Adapted with permission from ref. 91. Copyright 2018, American Chemical Society.)

Edge site catalytic activity can be further controlled by assembly on the support. Nørskov *et. al.* discovered that the support can induce significant changes in the hydrogen binding energy by long-range can der Waals (vdW) forces, and the HER performance were elevated[93]. Experimental result also show the strong elevated HER performance by metal support, Cao *et. al.* found that the interfacial interaction with the metal substrate can elevate hydrogen evolution performance[94].

**Amorphous $MoS_2$:**

Amorphous $MoS_2$ has been found to show enhanced catalytic performance than pristine $MoS_2$[95], for that in amorphous $MoS_2$, the Mo-Mo bond is much shorter and is pivotal to elevated HER performance[26]. Amorphous $MoS_3$ were fabricated by Jaramillo *et. al.* by wet chemical synthesis procedure, and during the electrocatalytic test, surface $MoS_3$ were transferred to amorphous $MoS_2$ to exhibit catalytic performance ~200 mV (10 mA $cm^{-2}$)[96]. Sang Chul Lee *et. al.* synthesis amorphous $MoS_2$ and explored the electrochemical HER performance with the TEM characterization[97]. They discovered that the amorphous $MoS_2$ can crystallize and deactivated during HER. As depicted in Figure 25, Phong D. Tran *et. al.* clarified the detailed catalytic route of amorphous $MoS_2$ by $[Mo_3S_{13}]^{2-}$ polymer contain four types of distinct potential catalytic site, and verified that molybdenum hydride moieties was responsible for the active HER catalytic site[98].

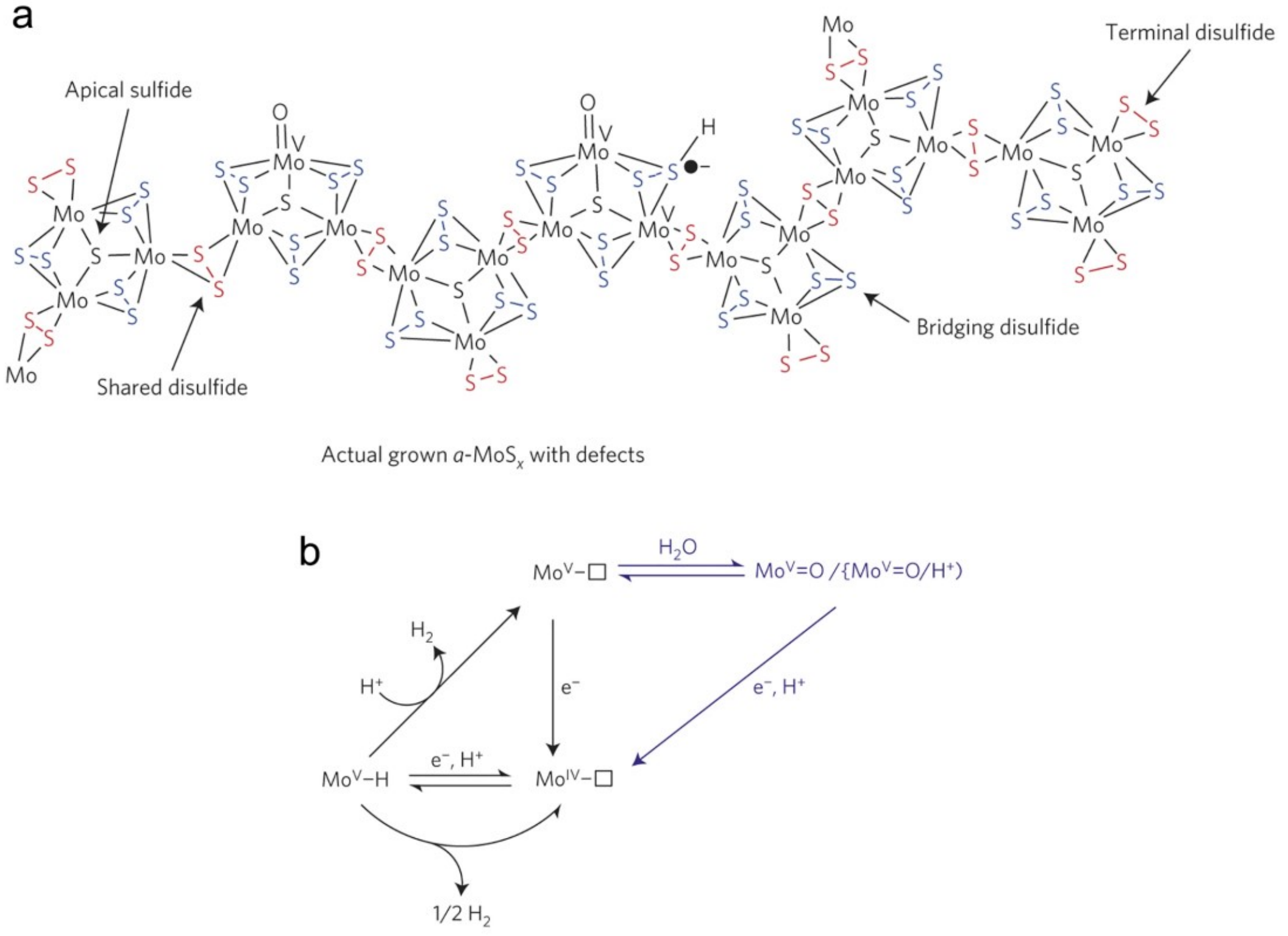

**Figure 25**. (a) Schematic structural for amorphous $MoS_2$ four different ligands: apical sulfide $\mu$-$S^{2-}$, bridging disulfides $(S–S)_{br}^{2-}$, shared $(S–S)_{sh}^{2-}$ and terminal disulfides $(S–S)_t^{2-}$ (b) Proposed catalytic pathway for $H_2$ evolution[98]. (Adapted with permission from ref. 98. Copyright 2016, Nature Publishing Group)

**Heterojunction：**

The heterojunction can elevate the HER performance since the favored chemisorption behavior[99], modification of the electronic states.

$MoS_2$ boundary of hetero 1T-2H phase were constructed, and the heterojunction display superior HER performance in both acidic and alkaline solution as well as long-term stability[99]. In their work,

J. Zhu *et. al.* made a detailed comparation of catalytic performance between 2H-2H domain and 2H-1T domain, and the theoretical results show that the 2H-1T boundary exhibit $\Delta G_{H*}$ = -0.13 eV (close to Pt (111) surface as well as Mo-edge of 2H-$MoS_2$). Long term test show that the boundary catalyst is stable. In-plane $MoS_2$ heterojunction has also been synthesized by Zili Wu *et. al.*, and find the similar results[100].

$MoS_2$ hybrid materials have been synthesized by forming heterojunction with carbon species, graphene, metal oxide, metal hydroxide, nitride, to elevate catalytic performance by synergistic effect.

$MoS_2$@$TiO_2$ heterojunction hybrid material has been synthesized and the hybrid materials owns better performance based on promoting the carrier transfer efficiency and prevent $MoS_2$'s aggregation[101]. $MoS_2$/$CoS_2$ were fabricated by sulfur assist pyrolysis of Co/Mo-MOF with 4~7 nm thick nanosheet, and the HER overpotential was 75 mV (10 mA $cm^{-2}$, 1 M KOH; commercial Pt/C's overpotential: 60 mV), which was much lower that $MoS_2$ (309 mV) or $CoS_2$ (525 mV)[102]. And $MoS_2$/$CoS_2$ shows an overpotential around 99 mV (10 mA $cm^{-2}$, 0.5 M $H_2SO_4$), 151 mV (10 mA cm-2, 0.5 M $Na_2SO_4$). $MoS_2$-$Ni_2O_3H$ were synthesized by two sequence hydrothermal reaction, the resultant hybrid catalyst show a overpotential of 84 mV at 10 mA $cm^{-2}$, 217 mV at 200 mA $cm^{-2}$ (1 M KOH)[103]. $Ni(OH)_2$/$MoS_2$ hybrid materials has been developed and display superior activity than Pt/C in alkaline HER[104]. The hybrid catalyst were formed by quantum 1T-$MoS_2$ decorated with $Ni(OH)_2$, exhibit overpotential of 57 mV (10 mA $cm^{-2}$) and 112 mV (100 mA $cm^{-2}$) in 1 M KOH. The intensified HER performance in alkaline can be attribute from $Ni(OH)_2$ that provide hydroxyl adsorption active sites, which facilitate the HER process. As depict in Figure 26, $MoS_2$ were firstly fabricated on carbon cloth, then $Ni(OH)_2$ nanoparticles were loaded on $MoS_2$ by electrodeposition[105]. Based on the synergistic interface effect of $Ni(OH)_2$/$MoS_2$, the HER performance were significantly elevated.

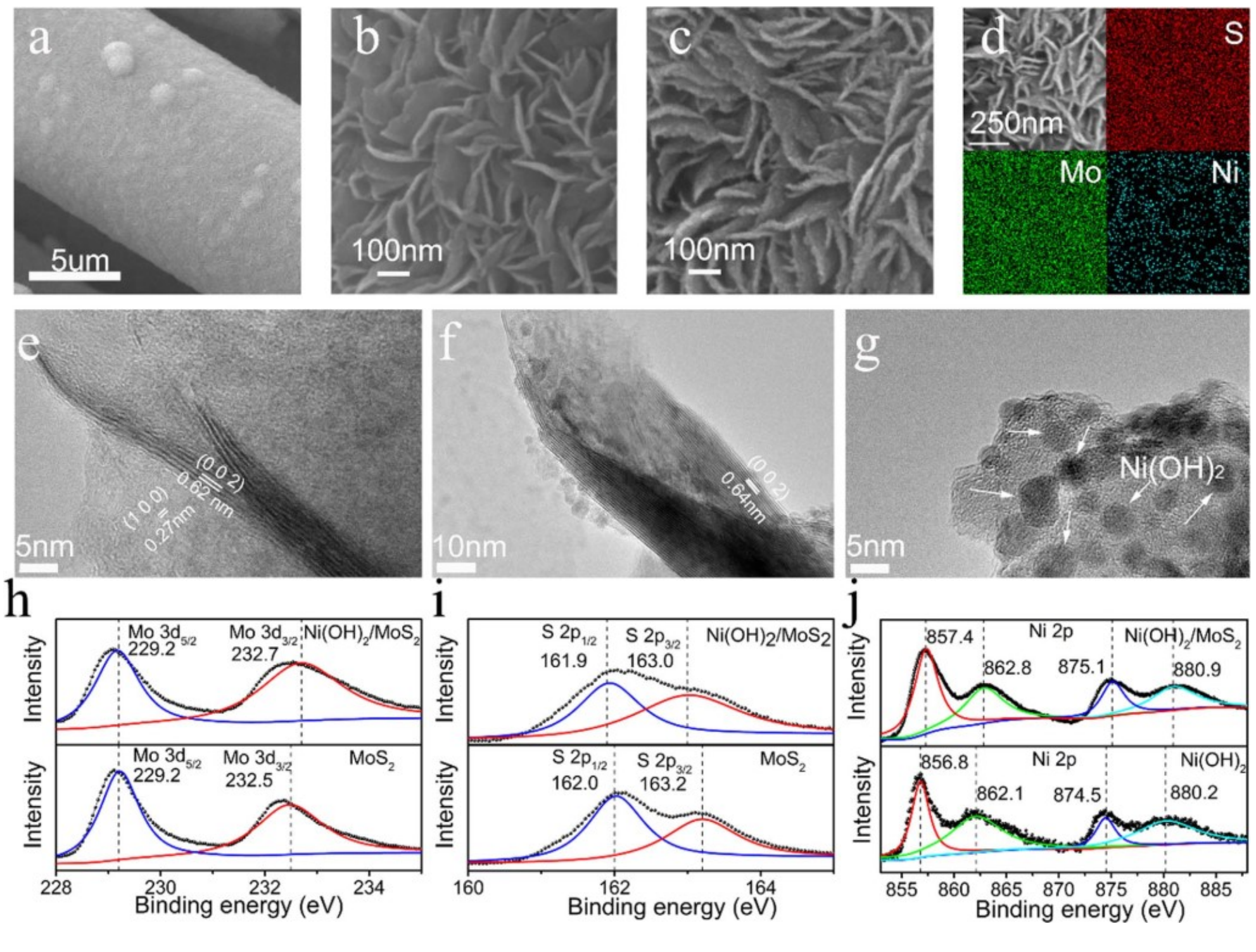

**Figure 26.** (a-b) SEM images of $MoS_2$@CC with different magnifications. (c-d)SEM images and EDX mappings of $Ni(OH)_2/MoS_2$@CC. (e) TEM image of MoS2@CC. (f-g) TEM image of $Ni(OH)_2/MoS_2$@CC. (h-j) XPS spectra of Mo 3d, S 2p and Ni 2p[105]. (Adapted with permission from ref. 105. Copyright 2017, Elsevier Ltd.)

Yuting Luo *et. al.* discovered $Co(OH)_2$ confined $MoS_2$ shows elevated HER performance in alkaline solution exhibiting with 15 mV overpotential[106] The hybrid materials are synthesized by two-step strategies. In the first step, $Li^+$ were introduced into the layer space of $MoS_2$ by the typical Butyllithium treatment method. Then in the second step, $Co^{2+}$ were exchanged into alkaline media to formulate $MoS_2$ intercalated with $Co(OH)_2$ nanoparticles. Yang *et. al*. fabricate 2H-$MoS_2$ and NiCo-LDH hybrid materials that exhibit a overpotential at 78 mV (10 mA $cm^{-2}$, 1 M KOH)[107].

As depicted in Figure 27, $MoS_2/CoSe_2$ hybrid catalyst were synthesized by Ming-Rui Gao *et. al.*, the hybrid material shows intensified HER performance[108]. The synthesis procedure was carried out with two steps. Firstly, CoSe were synthesized and stabilized with EDTA. Then the as-synthesized CoSe nanobelt were mixed with $(NH_4)_2MoS_4$ and undergo hydrothermal reaction to get the $MoS_2/CoSe_2$ hybrid catalyst. And the electrocatalysis performance is elevated when the catalyst undergoes the HER reaction.

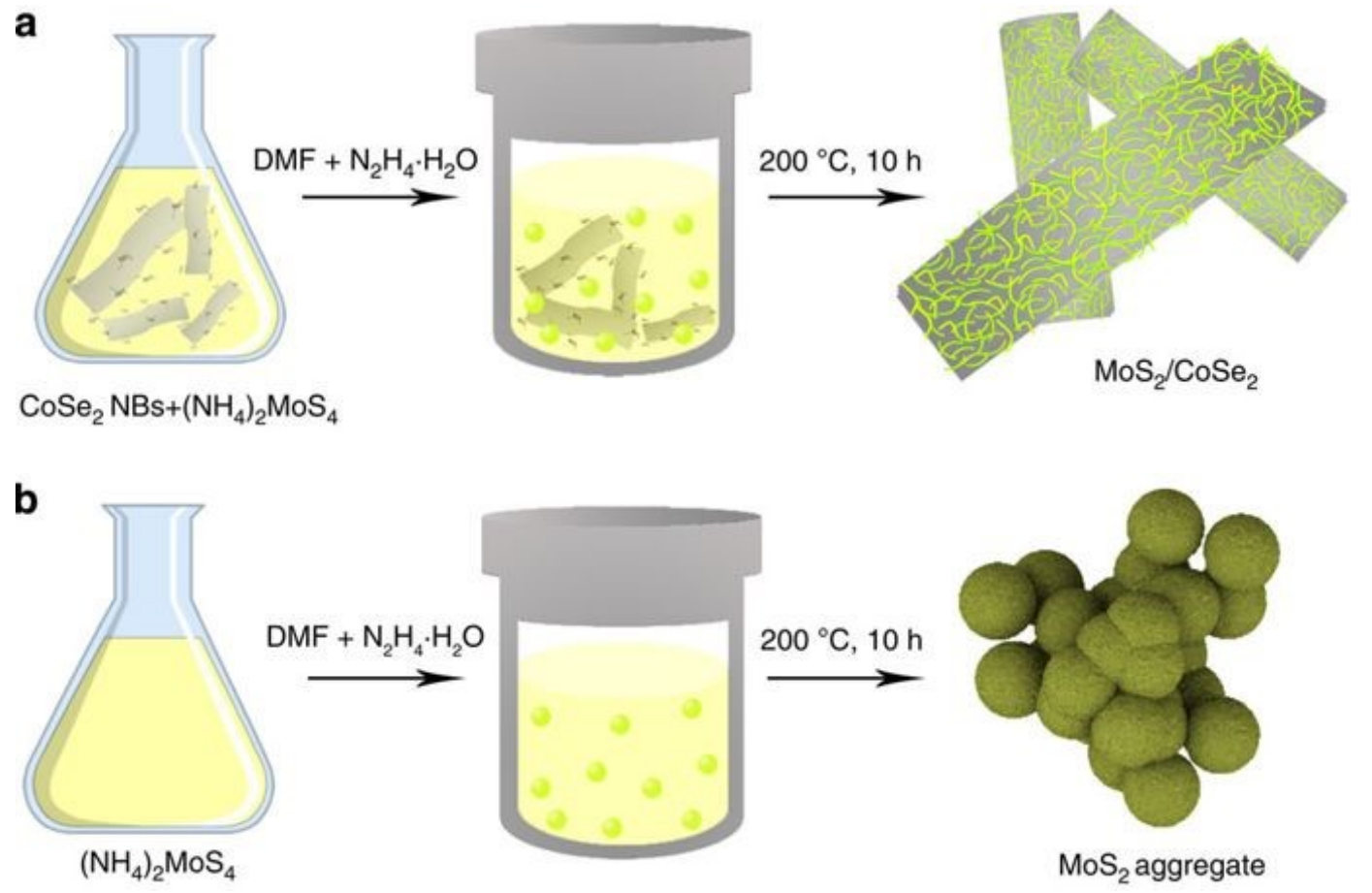

**Figure 27.** (a) Solvothermal synthesis with $CoSe_2$/DETA nanobelts as substrates for preparation of $MoS_2/CoSe_2$ hybrid. (b) Solvothermal synthesis without $CoSe_2$/DETA nanobelts leads to free $MoS_2$ nanosheet aggregates[108]. (Adapted with permission from ref. 108. Copyright 2015, Nature Publishing Group)

$MoS_2$ has also been loaded on $Ti_3C_2$ applying microwave heating with $(NH_4)_2MoS_4$ as precursor molecules. The author see that the $MoS_2$ is vertically aligned on $Ti_3C_2$ and edge active catalytic site is totally exposed[109]. Thin $MoS_2$ nanosheet (4~6 layers) were loaded on TiN nanorods, and the resultant hybrid catalyst exhibit a low overpotential of 119 mV (10 mA $cm^{-2}$), 44.8 (Tafel slope), and long term stability[110]. $MoS_2$-MoP heterojunction materials has been synthesized by Aiping Wu A et. al. by two-step synthesis procedure[111]. Firstly, hierarchical $MoS_2$ were synthesized and then as-synthesized $MoS_2$ were react with $NaH_2PO_2$ to get MoP coated $MoS_2$. The achieved hierarchical $MoS_2$@MoP core-shell catalyst HER performance exhibit overpotential of 42 mV (10 mV$cm^{-2}$ in 1M KOH), 29 mV (10 mV $cm^{-2}$ in 0.5M $H_2SO_4$). $Cu_7S_4$@$MoS_2$ hetero-nanocomposite show the HER performance with a 206 mV overpotential (0.5 M $H_2SO_4$, 200 mA $cm^{-2}$)[112]. And after CV activation for 11000 cycles with a Pt wire counter electrode, the catalytic performance is better than 20 % commercial Pt/C catalyst (overpotential: 26 mV in 10 mA $cm^{-2}$). The characterization show that during the CV activation, Pt atoms on Pt wire were dissolved but deposit on $Cu_7S_4$@$MoS_2$.

**Nano structural and Substrates engineering:**

Electrochemical process can be elevated by mass transport, electron transportation and diffusion, avoid of restacking, *etc*[113]. Constructing of various nanostructures are simply but efficient procedures for elevate the electrochemical performance[114]. Various structures like hierarchical structure[115], quantum dot[116], porous[117], hollow sphere[118], loading or on substrate[94, 119] were applied. As depicted in Figure 28 (a-b), assembling $MoS_2$ nanosheets into three-dimensional (3D) superstructure has been developed to access to maximum reactants and expose with more active site, high level electrochemical performance were achieved[115]. As depicted in Figure 28 (c), $MoS_2$ quantum dot (3.6 nm) were synthesized by one-step hydrothermal reaction and display with 160 mV overpotential with a 59 mV dec$^{-1}$ Tafel slope[116]. As depicted in Figure 28 (d), mesoporous $MoS_2$ with a double-gyroid (DG) morphology were synthesized by Thomas F. Jaramillo group, which preferentially expose abundant active edge sites on a large-area thin film, the mesoporous $MoS_2$ were formed by etching of the silica template[117]. As depicted in Figure 28 (e), hollow spheres were fabricated and elevated HER performance were achieved with a 112 mV onset overpotential ($\eta$=214 mV at 10 mA cm$^{-2}$)[118]. hydrothermal reaction between sodium molybdate, thioacetamide and oxalic acid at 200 °C for 24 h, and then calcite at 800 ºC in Ar atmosphere to form hollow structure and crystallization. As depicted in Figure 28 (f), $MoS_2$ were fabricated onto the carbon cloth vertically, which improve electron delivery effectively, and the HER performance were significantly elevated ($\eta$=205 mV at 200 mA cm$^{-2}$)[119]. As depicted in Figure 28 (g), $MoS_2$ were coupled to Ti substrate and forming interfacial tunneling barrier between Ti and $MoS_2$, which elevate HER performance that superior than Pt[94].

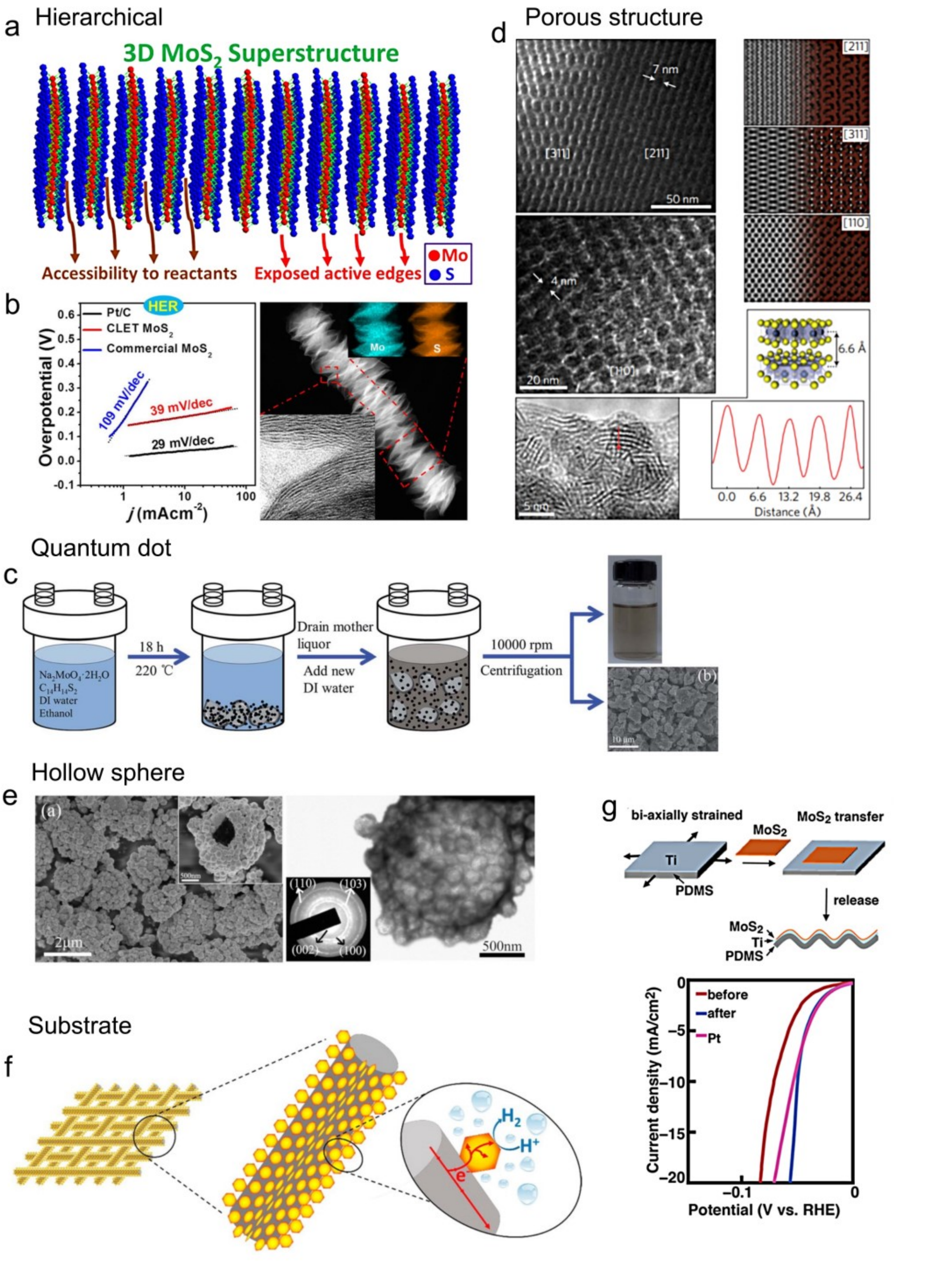

**Figure 28.** Nano structure and substrate fabrication of $MoS_2$ (a-b) Hierarchical structure of $MoS_2$ (a) diagram and (b) HER performance, TEM photo of 3D $MoS_2$ Superstructure with Maximized Accessibility to Reactants and Exposed Active Edges[115] (Adapted with permission from ref. 115. Copyright 2016, American Chemical Society.) (c) $MoS_2$ quantum dot. synthesis procedure to prepare $MoS_2$ QDs by using a hydrothermal approach[116] (Adapted with permission from ref. 116. Copyright 2016, American Chemical Society.) (d) porous $MoS_2$. Engineering the surface structure of $MoS_2$ to preferentially expose active edge sites for electrocatalysis[117]. (Adapted with permission from ref. 117. Copyright 2012, Nature Publishing Group) (e) $MoS_2$ hollow sphere. SEM and TEM spectra of micro-nano multi-hollow $MoS_2$[118] (Adapted with permission from ref. 118. Copyright 2016, American Chemical Society.) (f) Schematic diagram for synthesis of $MoS_2\perp$carbon cloth

substrate to expose as many $MoS_2$ edge sites[119] (Adapted with permission from ref. 119. Copyright 2015, American Chemical Society.) (g) Schematic illustration for the process of crumpling $MoS_2$ films and LSV curves shows better performance than Pt[94] (Adapted with permission from ref. 94. Copyright 2020, American Chemical Society.)

$MoS_2$'s layer is also correlated to the HER performance. $MoS_2$ with 1~4 layers were synthesized and the catalytic performance towards $MoS_2$ layer number were studied, it shows that the catalytic performance is consistent with the layer number[120].

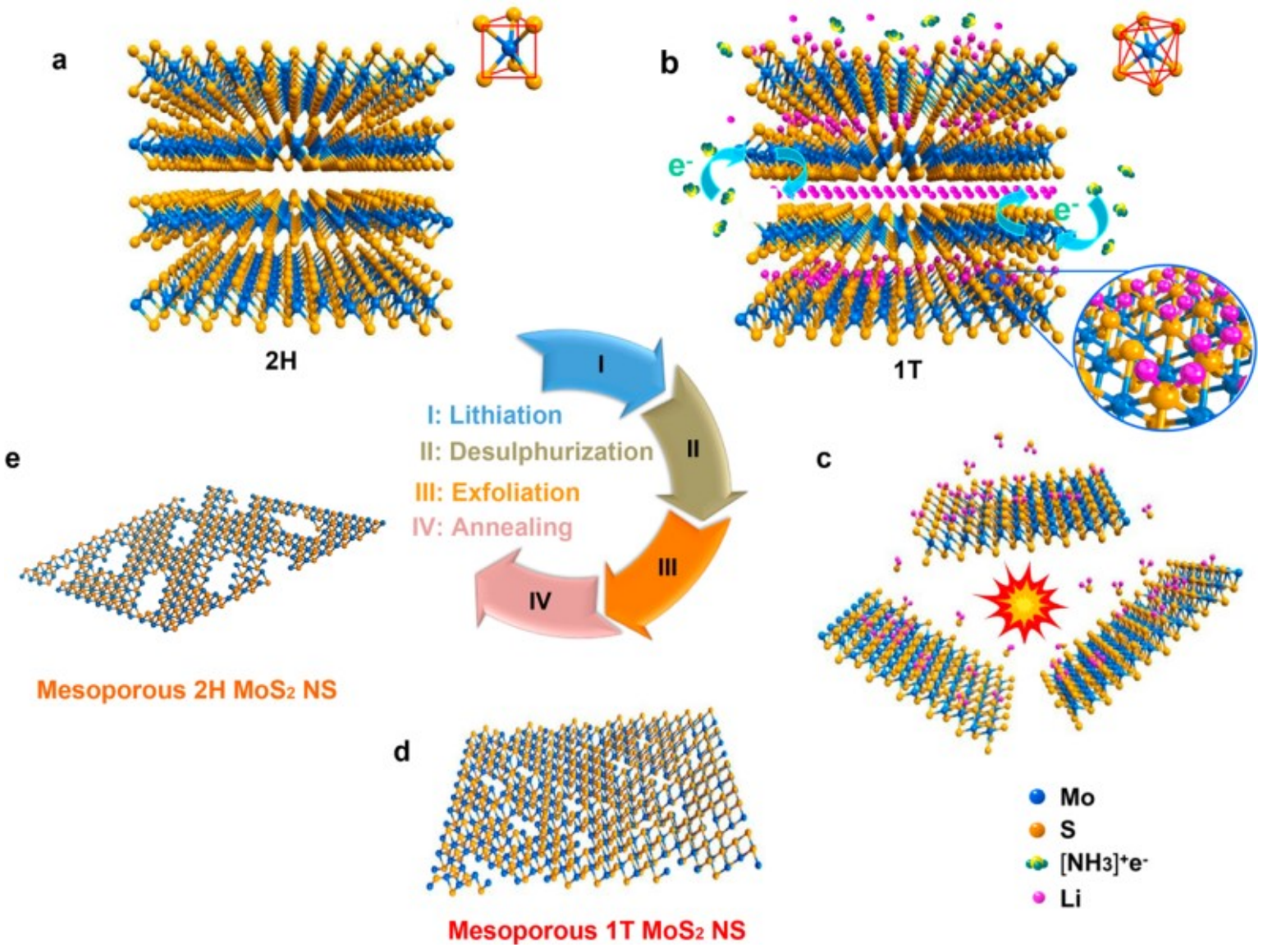

**Figure 29.** Schematic illustration of the preparation of mesoporous 1T phase $MoS_2$ nanosheets from bulk $MoS_2$ (a) by a LAAL process, including lithiation, desulfurization, and exfoliation (steps I, II, and III). Mesoporous 2H phase $MoS_2$ nanosheets (P-2H-$MoS_2$, e) can be obtained by a simple thermal annealing process from P-1T-MoS2 (step IV)[121] (Adapted with permission from ref. 121. Copyright 2016, American Chemical Society.)

As depicted in Figure 29, porous $MoS_2$ has been synthesized by chemical treatment of $MoS_2$ in the liquid $NH_3$ solution under low temperature by Song Jin *et. al.*, Mo and S atoms were etched by highly reactive $NH_3$ liquid to form nano hole[121]. The porosity $MoS_2$ expose more internal with edge catalytic active sites and facilitate the HER performance.
Carbon substrate support materials like carbon nanofibers[122-123], carbon cloth[119, 124], carbon paper[125], carbon nanotubes[126], graphene foam[127], graphene film[128], and Si/$SiO_2$[90, 129] have been applied in binding with $MoS_2$ to form heterojunction materials to exhibit strong synergistic effect[91] and improved conductivity[130]. And the porous nanostructure stabilized the electrode as well as the catalyst by timely release the yield $H_2$, which gives a much higher current density[131].

**Mechanism studies in $MoS_2$ water splitting, under-employed strategies for high performance HER catalysts**

Catalytic site

As depict in Figure 30 (a), $MoS_2$ were predicted to be an active electrocatalyst for water splitting based on theoretical calculation of thermochemical Gibbs energy changes in hydrogen evolution, the results show that adsorption energy is close to zero point which represent the benign adsorption and desorption process[132]. Then, in the year of 2007, Chorkendorff *et. al.* designed the controlling experiment and make a verification of the edge site catalytic site, they found that the HER performance increase with exposed edge atoms linearly (Figure 30 (b))[133]. Edge sites, sulfur vacancies, grain boundaries, alien catalytic sites were reaction sites for hydrogen evolution reactions, and Cao *et. al.* make a comparation of the intrinsic HER performance using TOF and Tafel slopes, which are 7.5 $s^{-1}$ (65~75 mV/dec) (edge site), 3.2 $s^{-1}$ (65~85 mV/dec) (sulfur vacancies), 0.1 $s^{-1}$ (120~160 mV/dec) (grain boundaries)[78]. Thomas F. Jaramillo *et. al.* made a detailed catalytic performance study on various intrinsic catalytic sites by TOF index[134]. And catalytic mechanism based on heterojunction engineering, interface engineering, were also studied. As shown in Figure (c), the catalytic mechanism of the sulfur vacancies and undercoordinated Mo were studied[80]. By systematic make samples with S defects with concentration ranging from 0~90 %, the HER performance prone to regulated by two rules, which corresponding to S vacancy and exposed Mo atom.

As depict in Figure (e-f), synergistic effect between $Ni(OH)_2$ and $MoS_2$ regulate the reaction route and reduced intermediate energy barrier, and exhibit elevated overall performance by adjust the H* intermediate adsorption energy $\Delta G$[105].

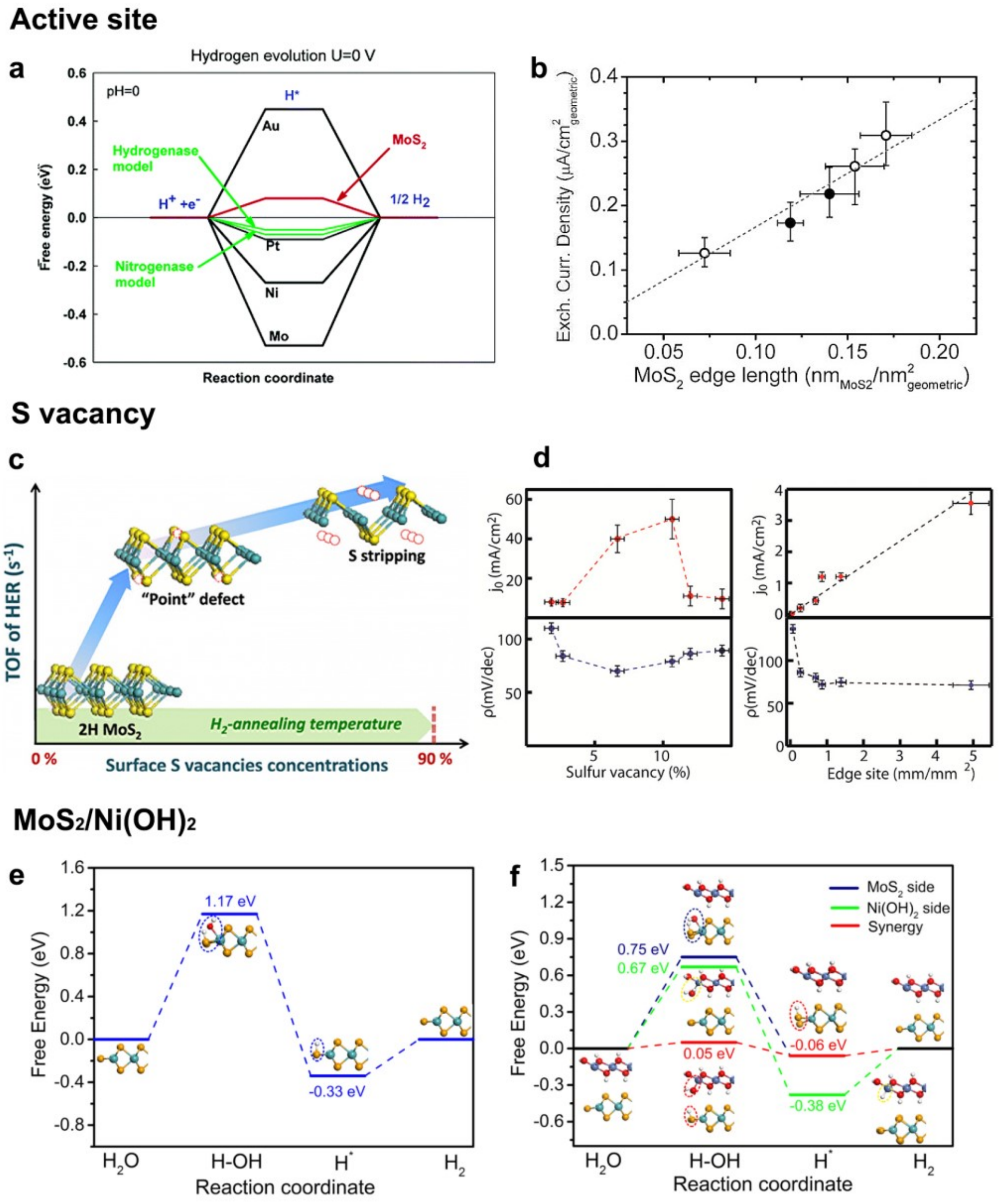

**Figure 30.** (a). Calculated free energy diagram for hydrogen evolution at a potential U = 0 (pH 0) [132]. (Adapted with permission from ref. 132. Copyright 2005, American Chemical Society.) (b) Corelate HER performance to $MoS_2$ edge site: Exchange current density versus $MoS_2$ edge length[133] (Adapted with permission from ref 133. Copyright 2007, American Association for the Advancement of Science) (c) Correlation between surface S vacancies and catalytic HER's TOF value[80] (Adapted with permission from ref. 80. Copyright 2019, American Chemical Society.) (d) Exchange current densities (upper) and Tafel slopes of the films as a function of the density of sulfur vacancies (left) and edge (right)[78] (Adapted with permission from ref. 78. Copyright 2016, American Chemical Society.) (e-f) DFT calculation of free energy diagram for HER by $MoS_2/Ni(OH)_2$ hybrid catalyst: free energy diagram for HER on the $MoS_2$ edge (e) and $Ni(OH)_2/MoS_2$ interface (f)[105] (Adapted with permission from ref. 105. Copyright 2017, Elsevier Ltd.).

Mechanism, Kinetic, and intermediate

To make a better understanding of the mechanism during HER, kinetics and intermediate studies were carried out. To study catalytic reaction on S vacancies, Kye Yeop Kim *et. al.* applied multiscale simulation combining *ab initio* calculations and kinetic Monte Carlo (KMC) simulations to examine HER reaction on $MoS_2$ monolayer S vacancy, the calculation result on TOF and polarization curve agrees well with experimental results, and the author discovered that hyper-reduced states in the HER reaction is playing an important role[135]. When S vacancy site is hyper-reduced with extra electrons by applying overpotential, HER barrier would decrease. Jiang *et. al.* make a study to focus on HER performance of 1T-$MoS_2$, they found that HER can readily take place on the basal plane by Volmer-Heyrovsky route, if Mn, Cr, Cu, Ni, Fe were doped in 1T-$MoS_2$, better performance can be arrived[136]. And the HER mechanism on “zigzag” and “armchair” 1T-2H interface were examined by DFT simulation, the result show that Volmer-Tafel is more energetically favorable route[137]. While another paper found that on the 1T′-2H interface, the Volmer-Heyrovsky route is more convenient[138]. Kinetic control[139-140] improvement on mass-transport is achieved by construct 2D channel to facilitate timely reactant supply and rapid gas release. Orbital control[65] is found by Qian *et. al.* to show that it will facilitate water adsorption behavior in C doped $MoS_2$. Looking forward to theoretical calculation, useful and even instructive information can be arrived[141]. $MoS_2$ nanocone arrays were fabricated by $SF_6/C_4F_8$ plasma assist etching, and the Tafel slope value were decreased with 50 mV $dec^{-1}$, the results is corresponding to kinetic optimize effect[142].
Key intermediate species and surface bonding structure changes of the catalysts can be identified by *in-situ* XAS characterization method, as shown in Figure 32 (b) that S-S in amorphous surface $Mo^{III}$-(S-S) were broken to bind to H and form in electrocatalysis procedure[143]. (Evidence from in Situ X-ray Absorption Spectroscopy for the Involvement of Terminal Disulfide in the Reduction of Protons by an Amorphous Molybdenum Sulfide Electrocatalyst).
Electrocatalysis is not just catalysis reactions, conductivity, adsorption and product release, catalytic site, electro-stability all play a role to the overall performance. To get better catalytic performance, taken two or more point into accounts is favorable to arrive better results[28].

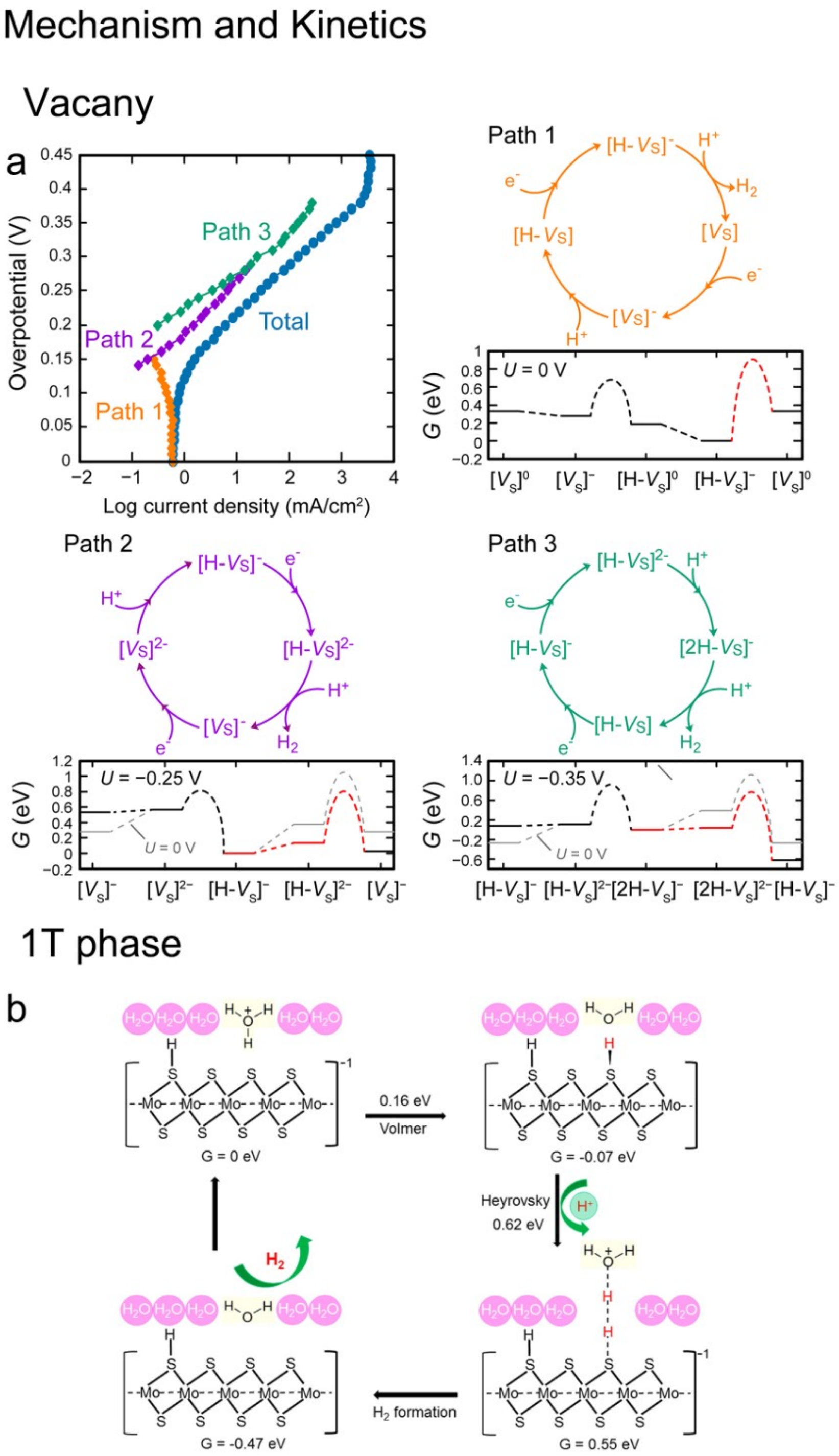

**Figure 31.** mechanism study on $MoS_2$ vacancies (a) and 1T phase $MoS_2$ (b). (a) Tafel plot of major paths in vacanciy catalytic site and the detailed pathways of Path 1–3 (The red line is the rate-limiting step and gray linesrepresent the corresponding path when the potential is 0). Paths 1–3 are the Tafel plots of major paths when overpotential is 0, –0.25, and −0.35 V, respectively[135]. (Adapted with permission from ref. 121. Copyright 2018, American Chemical Society.) (b) Overall reaction mechanism for HER on the surface of 1T $MoS_2$. Relative free energy (G) values and activation energies (for the Volmer and Heyrovsky steps) are also shown[136]. (Adapted with permission from ref. 135. Copyright 2016, American Chemical Society.)

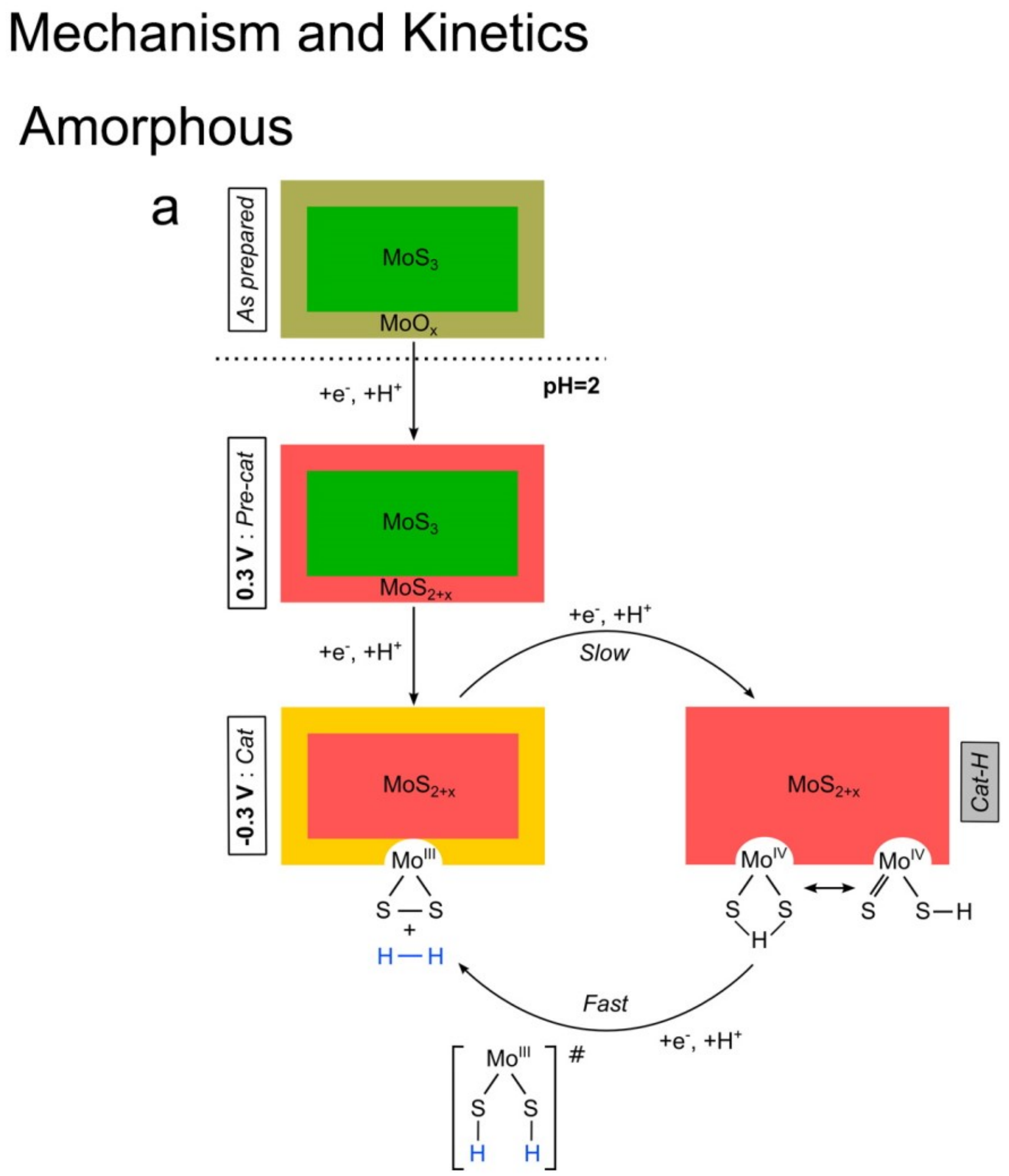

b

| sample | Mo K-edge EXAFS: number and nature of interaction | Mo K-edge EXAFS: $R$, Å | S K-edge XANES: sulfur oxidation state and binding mode | Mo L-edge XANES: formal oxidation states of molybdenum |
|---|---|---|---|---|
| dry (as prepared) | 1 Mo−O<br>3 Mo−S<br>1 Mo−Mo | 1.74<br>2.44<br>2.75 | $\mu$-$\eta^2$ $S^{2-}$, $\mu$-$\eta^3$ $S^{2-}$, $\mu$-$\eta^2$:$\eta^2$ $S_2^{2-}$ | Mo(V) or Mo(IV)[a] |
| ↓ | shortening of Mo−S bond, increased number of Mo−Mo bonds | | reduction | reduction |
| 0.3 V (precatalytic) | 6 Mo−S<br>1 Mo−Mo | 2.40<br>2.74 | $\mu$-$\eta^2$ $S^{2-}$, $\mu$-$\eta^3$ $S^{2-}$, $\mu$-$\eta^2$:$\eta^2$ $S_2^{2-}$ | Mo(IV) |
| ↓ | shortening of Mo−S bond, diminished Mo−Mo interactions | | oxidation | reduction |
| −0.3 V (catalytic) | 6 Mo−S | 2.36 | $\mu$-$\eta^2$ $S^{2-}$, $\mu$-$\eta^3$ $S^{2-}$, $\eta^2$ $S_2^{2-}$ | Mo(IV), Mo(III) |

[a]In $MoS_3$, the molybdenum ion is described as Mo(V) in the Hibble model and as Mo(IV) in the Weber model.

**Figure 32**. (a) Proposed Changes and Catalytic Cycle for the $MoS_x$ Film as Prepared and at pH = 2 under Pre-Catalytic and Catalytic Conditions (b) the corresponding Spectroscopic Features[143]. (Adapted with permission from ref. 143. Copyright 2015, American Chemical Society.)

**Table 1.** Typical superior Pt-like HER electrocatalysts

| Items | Synthesis Strategies | Synthesis Methods | HER activity | Reference |
|---|---|---|---|---|
| C-$MoS_2$ | Non metal doping | $Mo_2C$, 600 °C partial sulfurization | 45 mV at 10 mA $cm^{-2}$(1 M KOH) (close to Pt/C) | Nature Commun. 2019, 10, 1217 |
| NiO-1T $MoS_2$ | Co-doping | $[NiH_6Mo_6O_{24}]^{4-}$, thioacetamide, carbon fiber paper (180 °C, 24 h) | >80 mA $cm^{-2}$ (better than Pt/C) (1 M KOH) | Nature Commun. 2019, 10, 982 |
| $Ni_2P/MoS_2$/ N:RGO | Heterojunction, Substrate | $NiMoS_4$, N-RGO (RGO $NH_3$ treat at 800 °C), $NaH_2PO_2$ calcition | >200 mA $cm^{-2}$ (better than Pt/C) (0.5 M $H_2SO_4$) | Adv. Funct. Mater. 2019, 29, 1809151 |
| Co, Se co-doped $MoS_2$ | Co-doping, porous structure | $SiO_2$ template assist hydrothermal by $(NH_4)_6Mo_7O_{24}\cdot4H_2O$, $CS_2$, and Se powder | >208 mA $cm^{-2}$ (better than Pt/C, 0.5 M $H_2SO_4$) | Nature Commun. 2020, 11, 3315 |
| Co(0)/1T $MoS_2$ | Basal plane activation | Cobalt nanodisk and $MoS_2$ nanosheet interface reaction under sonication | 42 mV (10 mA $cm^{-2}$, 0.5 M $H_2SO_4$) Tafel slope 32 mV $dec^{-1}$ | Nature Commun. 2019, 10, 5231 |
| W-$MoS_2/MoO_2$ /CNT | Doping, porous substrate, | $(NH_4)_6Mo_7O_{24}\cdot4H_2O$, $(NH_4)_6H_2W_{12}O_{40}\cdot xH_2O$ load onto CNT, then calcite at 400 °C, sulfurization at 550 °C | 62 mV (10 mA $cm^{-2}$), Tafel slope 44 mV $dec^{-1}$ () | J. Mater. Chem. A, 2020,8, 14944-14954 |
| 1L $MoS_2$/Ti/PDMS | 7~10 % S vacancy, Ti Substrate effect (low interfacial tunneling barriers) | CVD synthesis of S vacancy-1L $MoS_2$, transfer to Ti substrate | Pt-like performance (0.5 M $H_2SO_4$), better than Pt/C (> ~7 mA $cm^{-2}$) | ACS Nano. 2020, 14, 1707-1714 |

**Conclusion:**

$MoS_2$ is a promising substitution HER catalyst for Platinum, for $MoS_2$ are abundant, non-toxic, and stable during the catalysis reaction. Nano scale engineering like doping, vacancy, defect, edge exposure, phase engineering strategies were demonstrated for adjustment of $MoS_2$'s intrinsic properties. What's more, heterojunction engineering, interfacial engineering, substrate effect have been developed for elevate $MoS_2$'s HER performance. Recently, more works have devoted to apply multiple synergistic strategies to achieve better catalytic performance and stability. Accompanied with above strategies, Pt-like activity or even superior activity can be achieved. The catalytic site and the mechanism for HER reaction were identified and clarified in detail by the theoretical as well as experimental work. Since a wealthy of various TMD materials, $MoS_2$ and other TMD catalyst with better performance is expected to achieved.